\documentclass[aps,prx,twocolumn,amsmath,amssymb,superscriptaddress,floatfix]{revtex4-2}

\usepackage{mathtools}
\usepackage{multirow}
\usepackage{xcolor}
\usepackage{bm}
\usepackage{amsfonts,amssymb,amsmath}
\usepackage{graphicx,dcolumn,bm,xcolor,braket,slashed}
\usepackage{times} 
\usepackage{comment}
\usepackage{array}
\usepackage{textcomp}
\usepackage[normalem]{ulem}
\renewcommand{\v}[1]{{\bf #1}}
\newcommand{\be}{\begin{eqnarray}}
\newcommand{\ee}{\end{eqnarray}}
\newcommand{\bbm}{\begin{bmatrix}}
\newcommand{\ebm}{\end{bmatrix}}
\newcommand{\bpm}{\begin{pmatrix}}
\newcommand{\epm}{\end{pmatrix}}

\begin{document}

\title{Revisiting the magnetic responses of bilayer graphene from the perspective of the quantum distance}

\author{Chang-geun Oh}
\email{cg.oh.0404@gmail.com}
\affiliation{Department of Applied Physics, The University of Tokyo, Tokyo 113-8656, Japan}
\author{Jun-Won Rhim}
\affiliation{Department of Physics, Ajou University, Suwon 16499, Republic of Korea}
\affiliation{Research Center for Novel Epitaxial Quantum Architectures, Department of Physics, Seoul National University, Seoul 08826, Republic of Korea}
\author{Bohm-Jung Yang}
\email{bjyang@snu.ac.kr}
\affiliation{Department of Physics and Astronomy, Seoul National University, Seoul 08826, Republic of Korea}
\affiliation{Center for Theoretical Physics (CTP), Seoul National University, Seoul 08826, Republic of Korea}
\affiliation{Institute of Applied Physics, Seoul National University, Seoul 08826, Republic of Korea}

\begin{abstract}
We study the influence of the quantum geometry on the magnetic responses of quadratic band crossing semimetals.
More explicitly, we examine the Landau levels, quantum Hall effect, and magnetic susceptibility of a general two-band Hamiltonian that has fixed isotropic quadratic band dispersion but with tunable quantum geometry, in which the interband coupling is fully characterized by the maximum quantum distance $d_\mathrm{max}$.
By continuously tuning $d_\mathrm{max}$ in the range of $0\leq d_\mathrm{max}\leq 1$, we investigate how the magnetic properties of the free electron model with $d_\mathrm{max}=0$ evolve into those of the bilayer graphene with $d_\mathrm{max}=1$.
We demonstrate that despite sharing the same energy dispersion $\epsilon(\bm{p}) =\pm\frac{p^2}{2m}$, the charge carriers in the free electron model and bilayer graphene exhibit entirely distinct Landau levels and quantum Hall responses due to the nontrivial quantum geometry of the wave functions.
\end{abstract}

\maketitle

\section{Introduction}
Graphene and its multilayer structures have garnered significant attention in various research fields, partly because of their unique carrier dynamics, sensitively dependent on the layer numbers~\cite{ando1998impurity, ando2005theory, katsnelson2006chiral,young2009quantum, mccann2006weak, tikhonenko2009transition, lu2019superconductors, liu2021nematic, cao2018unconventional, cao2018correlated, choi2021correlation,guerci2021moire}
. 
One notable example is the distinct quantum Hall effect (QHE) in monolayer and bilayer graphenes~\cite{novoselov2005two,zhang2005experimental, novoselov2006unconventional}. 
In monolayer graphene with linear band touching points, the carriers with pseudo-relativistic dispersion result in the half-integral quantum Hall plateaus characterized by the unique Landau Level spectrum with the energy $\epsilon^{\mathrm{single}}_N = \pm v_F\sqrt{2e\hbar B |N|}$, where $v_F$ is the Fermi velocity~\cite{novoselov2005two,zhang2005experimental,novoselov2006unconventional}, $N$ is an integer,  $B$ is a magnetic field, and $e$, $m$ are the electron's charge and mass, respectively.  
This unique behavior arises from the $\pi$ Berry's phase of the Dirac points. 

\begin{figure}[t]
\includegraphics[width=85mm]{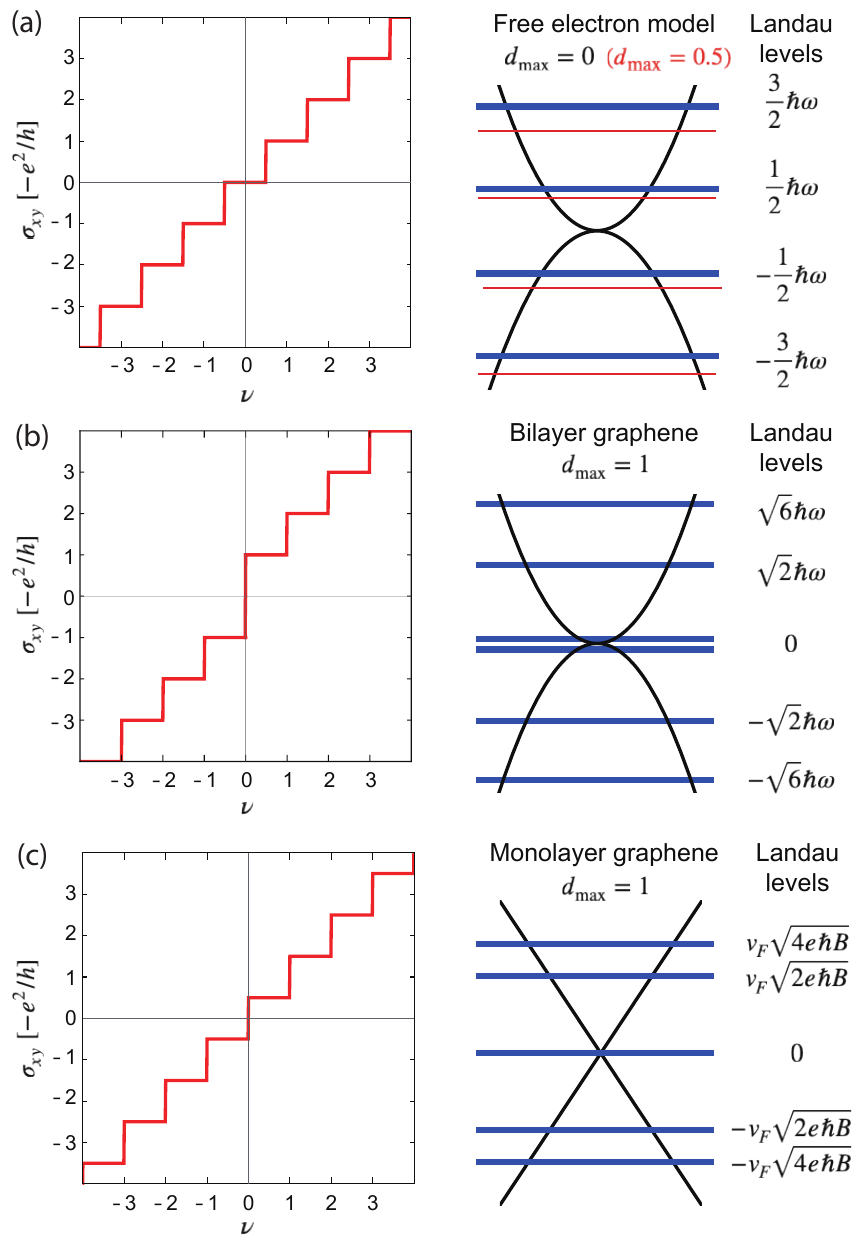} 
\caption{\label{fig1}
Schematics of the representative integer quantum Hall effect (left) and the corresponding Landau level spectrum (right) for 
	(a) free electron model, characterized by geometric triviality ($d_\mathrm{max}=0$), 
	(b) bilayer graphene, featuring nontrivial geometry with $d_\mathrm{max}=1$, 
	and (c) monolayer graphene, also exhibiting geometric nontriviality with $d_\mathrm{max}=1$. 
	The red horizontal lines in (a) represent the Landau levels of a quadratic band crossing in Eq.~(\ref{eq:Ham}) featuring nontrivial quantum geometry with $d_\mathrm{max}=0.5$.
}
\end{figure}

On the other hand, in Bernal stacked bilayer graphene, simply bilayer graphene hereafter,
the charge carriers exhibit a parabolic dispersion $\epsilon(p) =\pm\frac{p^2}{2m}$ with the effective mass $m$, hosting the Landau levels $\epsilon^{\mathrm{bilayer}}_N=\pm \hbar\omega\sqrt{N(N-1)}$. In bilayer graphene, the zero energy Hall plateau is absent, which is attributed to the $2\pi$ Berry phase around the quadratic band crossing point~\cite{novoselov2006unconventional,fal2008electronic}. 
It is noteworthy that although the parabolic band dispersion of the bilayer graphene is identical to that of the free electron gas, these two systems display entirely different Landau levels and QHEs. 
This indicates the importance of considering not only the band dispersion but also the geometry of wave functions to correctly describe the magnetic responses.
The distinct Landau levels and QHEs of monolayer and bilayer graphenes are compared in Figure~\ref{fig1}, in which we also plot the free electron gas Landau levels $\epsilon^{\mathrm{conv}}_N=\pm\hbar\omega(N+\frac{1}{2})$ with the cyclotron frequency $\omega=eB/m$~\cite{girvin1987quantum,macdonald1989quantum}.

Recent studies have shown that the quantum distance is a central quantity characterizing the geometric properties of two-dimensional quadratic band touching systems, leading to intriguing phenomena such as anomalous Landau level spreading~\cite{rhim2020quantum,rhim2021singular} and the emergence of boundary modes in flat band systems~\cite{oh2022bulk,kim2023general}.
More explicitly, the Hilbert-Schmidt quantum distance, or simply quantum distance, in momentum space is defined as 
\begin{eqnarray}
d_{\mathrm{HS},n}^2(\bm{k,k'})=1-|\braket{\psi_{n,\bm{k}}|\psi_{n,\bm{k'}}}|^2, 
\end{eqnarray}
where $n$ is a band index and $\psi_{n,\bm{k}}$ is the Bloch eigenstate of the $n$-th band with crystal momentum $\bm{k}$.
In particular, it was shown that the maximum quantum distance, denoted as $d_\mathrm{max}$, determines various geometric properties of the quadratic band crossing semimetals, including the Berry's phase~\cite{hwang2021wave,jung2024quantum}.
In terms of $d_\mathrm{max}$, the geometry of bilayer graphene is characterized by $d_\mathrm{max}=1$, while the free electron with a quadratic band crossing is described by $d_\mathrm{max}=0$, indicating geometric triviality. 
Although these two systems have two distinct $d_\mathrm{max}$ values, to properly understand the role of the interband coupling in their distinct magnetic responses, one can design a model Hamiltonian that has fixed isotropic quadratic band dispersion $\epsilon(p) =\pm\frac{p^2}{2m}$ but with tunable quantum geometry. 

In this paper, we examine the Landau levels, QHE, and magnetic response functions of the geometrically generalized model. By thoroughly examining them, we illustrate the nontrivial role of the interband coupling measured by $d_\mathrm{max}$ in magnetic responses of quadratic band crossing semimetals.

The rest of the paper is organized as follows. 
In Sec.~\ref{sec:model}, we construct a model Hamiltonian for isotropic quadratic band touching semimetals, where the band dispersion remains $\pm p^2/(2m)$ while the geometry of wave functions is tunable. 
In Sec.~\ref{sec:LL}, we analyze the Landau levels of the model and examines the role of wave function geometry. 
In Sec.~\ref{sec:QHE}, we investigate the evolution of QHE between free electron gas and bilayer graphene. 
In Sec.~\ref{sec:MRF}, we further explore the influence of wave function geometry on magnetic response functions using the Roth-Gaou-Niu relation \cite{gao2017zero,roth1966semiclassical,fuchs2018landau}. 
Our concluding remarks can be found in Sec.~\ref{sec:conclusion}.
Appendixes contains the detailed calculations of Landau levels and a lattice model analysis.

\section{Model \label{sec:model}}
Let us construct a general Hamiltonian whose energy eigenvalues are given by 
\begin{eqnarray}
    \epsilon_\pm(\bm{k}) = \pm\frac{1}{2}(k_x^2+k_y^2). \label{eq:energy}
\end{eqnarray}
Explicitly, we consider the Hamiltonian
    \begin{eqnarray}
        \mathcal{H}_{0}({\bm{k}}) = \sum_{\alpha } h_\alpha ({\bm{k}}) \sigma_\alpha , \label{eq:Ham}
    \end{eqnarray}
where $\sigma_\alpha$ represents an identity ($\alpha=0$) and Pauli matrices ($\alpha = x,y,z$), respectively.
$h_{x,y,z} ({\bm{k}})$ are real quadratic functions given by $h_{z} ({\bm{k}}) = {-d\sqrt{1-d^2}} k_y^2,~h_{y} ({\bm{k}}) = d k_x k_y,~h_{x} ({\bm{k}}) = k_x^2/2+(1-2d^2)k_y^2/2$, and $h_{0} ({\bm{k}}) = 0$. 
Here, the parameter $d$ is defined as $d=\xi d_\mathrm{max}$ in which $\xi =\pm 1$, and $d_\mathrm{max}$ is the maximum value of the quantum distance $d_{\mathrm{HS},n}(\bm{k,k'})$ between all the possible pairs of wave functions at $\mathbf{k}$ and $\mathbf{k}^\prime$ with the given band index $n=1,~2$. 
The energy eigenvalues of $\mathcal{H}_{0}({\bm{k}})$ remain unchanged from Eq.~(\ref{eq:energy}) regardless of the value of \( d \) within the range \( -1 \leq d \leq 1 \).
When $d_\mathrm{max}=1$, the Hamiltonian in Eq.~(\ref{eq:Ham}) corresponds to the low energy Hamiltonian of the Bernal stacked bilayer graphene \cite{koshino2007orbital,mccann2013electronic}, and $\xi=\pm1$ is related to the valley index.
The parameters $d_\mathrm{max}$ and $\xi$ determine the Berry phase $\Phi_B$ \cite{hwang2021wave} and quantum geometric tensor $g_{ij}^n$ given by 
$g_{ij}^n=2\braket{\partial_{k_i} u_n(\bm k)| \partial_{k_j}u_n(\bm k)}-2\braket{\partial_{k_i} u_n(\bm k)| u_n(\bm k)}\braket{u_n(\bm k)|\partial_{k_j}  u_n(\bm k)}$, where $u_n(\bm k)$ is the $n$-th Bloch wave function~\cite{provost1980riemannian}.
Explicitly, for $\mathcal{H}_{0}({\bm{k}})$, the Berry phase and components of the quantum geometric tensor are given by
\begin{eqnarray}
    &&\Phi_B = -2\pi\xi\sqrt{1-d_\mathrm{max}^2}~~~~~(\mathrm{mod} 2\pi),   \label{eq:BP}\\
    &&g^n_{xx}(\bm{k})=2d_\mathrm{max}^2\frac{k_y^2}{k^4}, ~~g^n_{yy}(\bm{k})=2d_\mathrm{max}^2\frac{k_x^2}{k^4},\nonumber \\
    &&g^n_{xy}(\bm{k})=g^n_{yx}(\bm{k})=-2d_\mathrm{max}^2\frac{k_xk_y}{k^4}. \label{eq:QGT}
\end{eqnarray}
We note that more general quadratic band touching Hamiltonians, which exhibit anisotropic energy dispersions, require other geometric quantities to describe all possible interband coupling terms \cite{jung2024quantum}. However, when the system has rotational symmetry, a single geometric parameter $d_\mathrm{max}$ suffices to fully characterize the quantum geometry \cite{hwang2021wave,jung2024quantum}.

\begin{figure}[t]
\includegraphics[width=85mm]{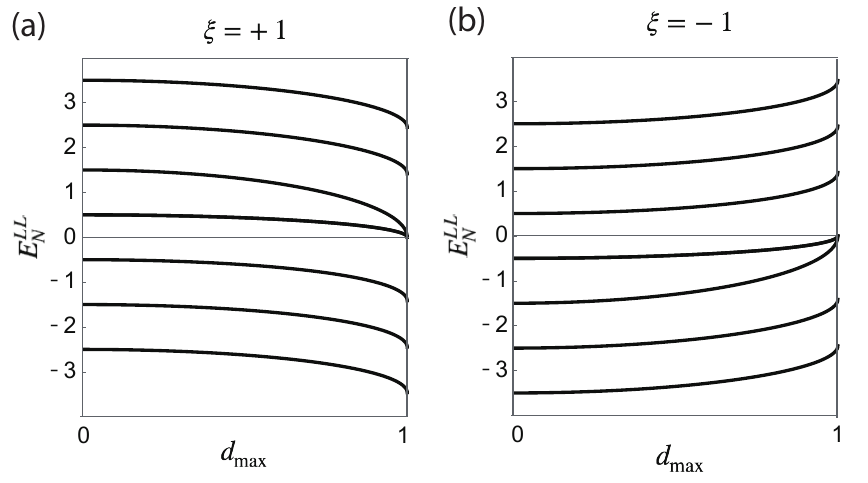} 
\caption{\label{fig2}
The evolution of the Landau levels $E_N^{LL}$ as a function of the maximum quantum distance $d_\mathrm{max}$ 
for (a) $\xi=+1$ and (b) $\xi=-1$, respectively, with $\hbar=\omega=1$.
} 
\end{figure}

\section{Landau levels \label{sec:LL}}
To understand the evolution of the Landau levels between $\epsilon_N^{conv}$ [Fig.~\ref{fig1}(a)] and $\epsilon_N^{bilayer}$ [Fig.~\ref{fig1}(b)], we introduce a {perpendicular} magnetic field $B$ to the Hamiltonian in Eq.~(\ref{eq:Ham}). 
We {consider the Landau gauge $\bm{A}=(0,Bx)$, which preserves translational invariance in the $y$-direction, and} replace the momentum by ladder operators as $k_x \to (a+a^\dagger)/(\sqrt{2}l_B)$ and $k_y \to i(a-a^\dagger)/(\sqrt{2}l_B)$, where $l_B=\sqrt{\hbar/eB}$ is the magnetic length, and $a$($a^\dagger$) is the annihilation (creation) operator.
To ensure the Hamiltonian's hermiticity, we perform symmetrization: $k_xk_y=(k_x k_y+k_y k_x)/2=i(a^2-(a^\dagger)^2)/(2l_B^2)$.
{Then, the Hamiltonian reads}
\begin{align}
H_{LL}=\frac{1}{2l_B^2}
\begin{pmatrix}
g_{11}&g_{12}\\
g_{21}&g_{22}
\end{pmatrix},
\end{align}
{where $g_{11}=-g_{22}=d\sqrt{1-d^2} (2 a^\dagger a+1 - a^{\dagger2} - a^2)$ and $g_{12}=g_{21}^\dagger= -(2a^\dagger a+1) + (a^2 - a^{\dagger2})d+(2 a^\dagger a +1-a^2-a^{\dagger2})d^2$. 
It can be verified that when $d_\mathrm{max} = 1$, this Hamiltonian corresponds to the low-energy effective Hamiltonian of bilayer graphene \cite{mccann2013electronic,mccann2006landau}. In contrast, when $d_\mathrm{max} = 0$, it corresponds to the conventional Hamiltonian, where electrons follow cyclotron orbits with the conventional Landau levels $\epsilon^{\mathrm{conv}}_N = \pm\hbar\omega(N + \frac{1}{2})$ \cite{landau1930diamagnetismus}. The case where $d_\mathrm{max}$ is not an integer has not yet been explored. By investigating the regime $0<d_\mathrm{max}<1$, one can study how the magnetic properties of the conventional model gradually evolve into those of bilayer graphene.}

Solving the transformed Hamiltonian yields the Landau levels (see Appendix for details):
\begin{eqnarray}
&&E^{LL}_0(d_\mathrm{max},\xi)= \frac{\xi}{2} \hbar \omega \sqrt{1-d_\mathrm{max}^2}, \nonumber \\
&&E^{LL}_1(d_\mathrm{max},\xi)=\frac{3\xi}{2}\hbar \omega \sqrt{1-d_\mathrm{max}^2}, \nonumber \\
&&E^{LL}_{N}(d_\mathrm{max},\xi)= \xi\hbar \omega \bigg(\sqrt{1-d_\mathrm{max}^2} \nonumber \\
 &&~~~~~~~~~~~~~~~~~~~+ \frac{\mathrm{sgn}(N)}{2}\sqrt{(2|N|-1)^2-d_\mathrm{max}^2}\bigg), \label{eq:LLs}
\end{eqnarray}
where $N = \pm2, \pm3, ...$ and $\mathrm{sgn}(N)$ represents the sign of $N$.

In Figure~\ref{fig2}, the $d_\mathrm{max}$-dependence of Landau levels is depicted. When $d_\mathrm{max}=0$, the Landau levels $E^{LL}_N$ are equivalent to $\epsilon_N^{conv}$ 
but start to deviate from $\epsilon_N^{conv}$ as $d_\mathrm{max}$ increases. When $d_\mathrm{max}$ reaches one, they become $\epsilon_N^{bilayer}$. The degeneracy of Landau levels between $E^{LL}_0(d_\mathrm{max}=1)$ and $E^{LL}_1(d_\mathrm{max}=1)$ leads to the absence of a zero energy plateau in QHE as described in Section~\ref{sec:QHE}.
Since such degeneracy of Landau levels occurs only when $d_\mathrm{max}=1$, the absence of the zero energy plateau cannot be observed if $d_\mathrm{max} \neq 1$.
%

One can verify that the degeneracy at $d_\mathrm{max}=1$ exists for both $\xi =\pm 1$. However, depending on $\xi$, the origin of zero Landau levels is different. For $\xi=+1$, the two zero energy levels come from the upper band, while for $\xi=-1$, the two zero energy levels come from the lower band. 
In a previous work~\cite{koshino2010anomalous}, this $\xi$-dependence of zero energy Landau levels was demonstrated by creating a gap between two bands in bilayer graphene. Here, on the other hand, we verify that the $\xi$-dependence of zero energy Landau levels by continuously varying the quantum distance.

Furthermore, when $d_\mathrm{max}=1$ or $d_\mathrm{max}=0$, the Landau levels are symmetric with respect to $E=0$, as shown in Figs.~\ref{fig2}(a) and (b). This result arises from chiral symmetry, represented by the operator $\sigma_z$, which satisfies $\sigma_z H_0(\bm{k})\sigma_z=-H_0(\bm{k})$, exclusively when $d_\mathrm{max}=1$ or $d_\mathrm{max}=0$. 
This symmetry holds even in the presence of a magnetic field (See Appendix). 
In fact, the chiral symmetry is crucial for the degeneracy observed at $d_\mathrm{max}=1$. 
To confirm this idea,  we introduce a perturbation  $H_{\text{pert}}=\delta k^2 \sigma_0$ that breaks the chiral symmetry, and subsequently calculate the resulting Landau levels. With this perturbation, the zeroth and first Landau levels for $d_\mathrm{max}=1$ shift to $E^{LL}_0= \frac{\xi}{2} \hbar \omega \delta$ and  $E^{LL}_1= \frac{3\xi}{2} \hbar \omega \delta$, respectively, thereby lifting the degeneracy. This demonstrates that the presence of the zero energy plateau necessitates chiral symmetry as well as $d_\mathrm{max}=1$. 

In addition, when the Hamiltonian possessses chiral symmetry, one can define a winding number ($W$): 
\begin{eqnarray}
W\equiv \int_C \frac{d\bm{k}}{2\pi}\left[ \frac{h_x}{|\bm{h}|} \nabla \left (\frac{h_y}{|\bm{h}|}\right)-\frac{h_y}{|\bm{h}|} \nabla \left(\frac{h_x}{|\bm{h}|}\right)\right].
\end{eqnarray}
Explicitly, for $d_\mathrm{max}=0$ and $d_\mathrm{max}=1$, we obtain
\begin{eqnarray}
&&W=0 ~~~~~\text{for}~~ d_\mathrm{max}=0 ~~\text{and} ~~\xi=\pm1,\nonumber\\
&&W=+2 ~~\text{for} ~~d_\mathrm{max}=1~~\text{and}~~ \xi=+1, \nonumber\\
&&W=-2 ~~\text{for} ~~d_\mathrm{max}=1~~ \text{and}~~ \xi=-1.
\end{eqnarray}
This indicates that the presence of the zero energy plateau is contingent upon chiral symmetry with a winding number of two.

For $0<d_\mathrm{max}<1$, the chiral symmetry is broken because the Hamiltonian in Eq.~(\ref{eq:Ham}) has nonzero $h_x$, $h_y$ and $h_z$, simultaneously. Consequently, the Landau levels are no longer symmetric with respect to $E=0$.
However, one can still find the symmetry between $\xi=+1$ and $\xi = -1$ cases in which the Landau levels have the opposite signs, as shown in Eq.~(\ref{eq:LLs}), Figs.~\ref{fig2}(a) and (b). For $|N|\gg1$, this can be understood using the semiclassical results given by \cite{roth1966semiclassical} 
\begin{eqnarray}
E_N = \hbar \omega (N+\frac{1}{2}-\frac{\Phi_B}{2\pi}), \label{eq:Onsa}
\end{eqnarray}
where $\Phi_B$ is Berry phase. For both $\xi=\pm1$ cases with $|N|\gg1$, the Landau levels in Eq.~($\ref{eq:LLs}$) are identical to those in Eq.~(\ref{eq:Onsa}).
Depending on $\xi$, the sign of $\Phi_B$ changes oppositely, as shown in Eq.~(\ref{eq:BP}), explaining the symmetric structure of the Landau levels between $\xi=+1$ and $\xi=-1$ cases. Interestingly, this symmetric structure persists even for low $N$ as explicitly shown in Eq.~(\ref{eq:LLs}).

\section{Quantum Hall effect \label{sec:QHE}}
The dependence of the Landau levels on $d_\mathrm{max}$ significantly influences the QHE. Here, we focus on the case $\xi=+1$; the results for $\xi=-1$ can be obtained by reversing the sign of the energies for $\xi=1$ as shown in Fig.~\ref{fig2}.
To understand the influence of $d_\mathrm{max}$ on QHE, it is more insightful to examine the magnetic field dependence of Hall conductivity or Hall resistivity rather than the filling factor dependence of Hall conductivity. This is because the continuous variation from $d_\mathrm{max}=0$ to $d_\mathrm{max}=1$ is not apparent in the latter case; instead, a sudden change is observed at $d_\mathrm{max}=1$ due to the zero energy degeneracy. Therefore, we analyze the magnetic field dependence of QHE by varying $d_\mathrm{max}$ when the electron density is fixed.
%
\begin{figure}[t]
\includegraphics[width=84mm]{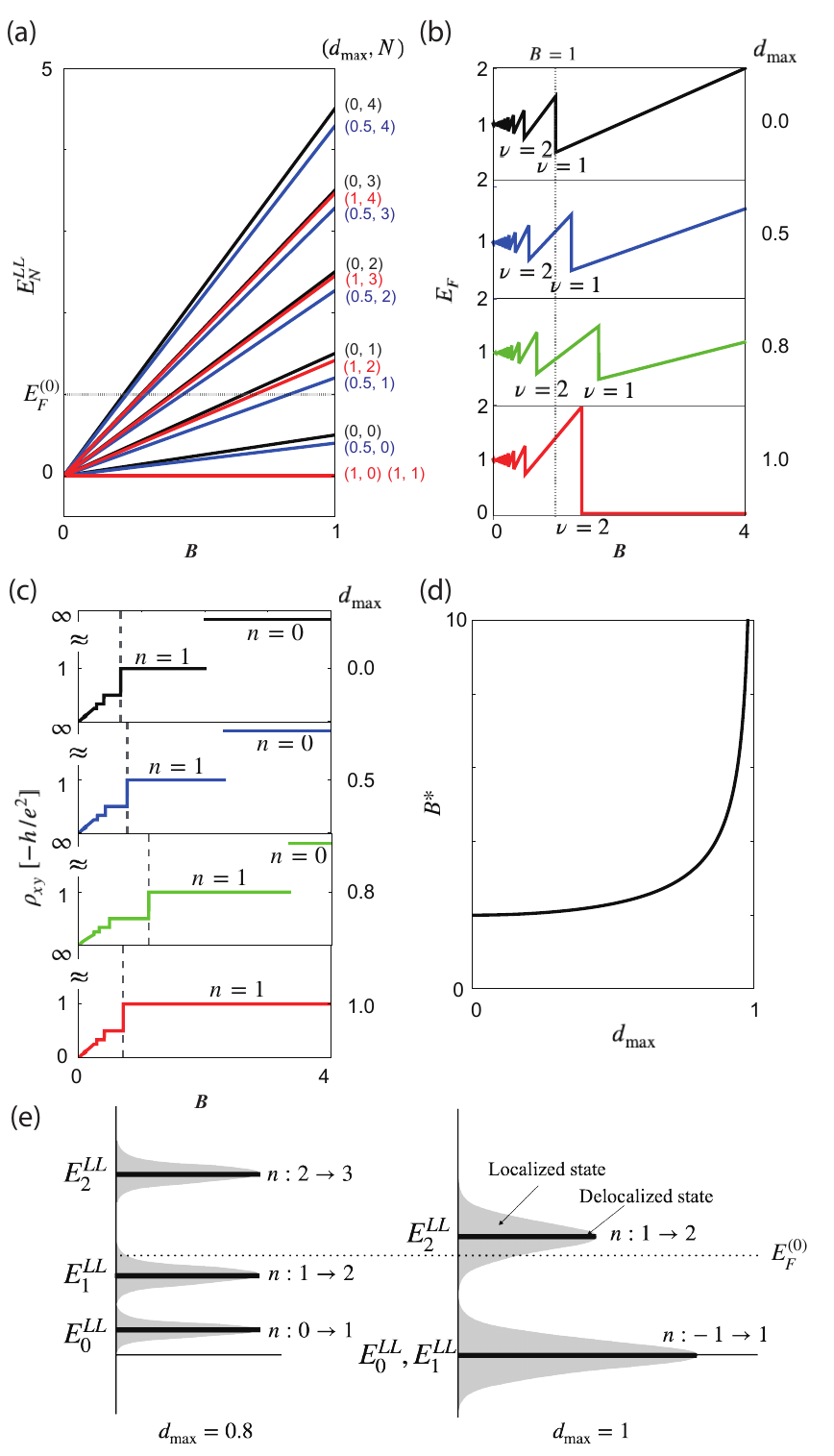} 
\caption{\label{fig3}
(a) The magnetic field $B$ dependence of Landau levels $E_N^{LL}$ for $d_\mathrm{max}=0, 0.5$ and $1$ with $\hbar=e=1$ and $\xi=+1$.
(b) $B$ dependence of the Fermi level $E_F$ for $d_\mathrm{max}=0, 0.5, 0.8$ and 1. (c) $B$ dependence of the Hall resistivity $\rho_{xy}$ for $d_\mathrm{max}=0, 0.5, 0.8$ and 1. 
The black, blue, green and red lines in (a-c) represent $d_\mathrm{max}=0, 0.5, 0.8$ and 1, respectively. 
(d) The maximum quantum distance $d_\mathrm{max}$ dependence of the magnetic field $B^*$, where the jump from $n=1$ to $n=0$ in $\rho_{xy}$ occurs. Here, we set the Fermi energy at zero magnetic field as $E_F^{(0)} = 1$.
(e) Schematics of Landau levels in disordered system for $d_\mathrm{max}=0.8$ and 1. The grey and black areas represent localized and delocalized states, respectively. $n: i \to i+1 (i=0,1)$ next to delocalized states indicate the corresponding jump in $\rho_{xy}$. Here, we consider the situation where Hall resistivity is on the $n=2$ ($n=1$) plateau for $d_\mathrm{max}=0.8$ ($d_\mathrm{max}=1$).
}
\end{figure}

Figure~\ref{fig3}(a) shows the magnetic field dependence of the Landau levels in Eq.~(\ref{eq:LLs}).
As the magnetic field increases, the topmost occupied level changes when the filling factor $\nu$ reaches integer values.
At theses points, $E_F$ transitions from $E^{LL}_{\nu}$ to $E^{LL}_{\nu-1}$.
Figure~\ref{fig3}(b) shows the magnetic field dependence of the Fermi energy $E_F(B)$, referred to as Shubnikov-de Hass oscillation, for various values of $d_\mathrm{max}=0, 0.5, 0.8$ and 1.
Increasing $d_\mathrm{max}$ shifts the magnetic field $B_{\nu=i}$ where the transition occurs with an integer $i$.
Furthermore, the slope of the Fermi energy at $B_{\nu=i}$, defined as 
\begin{eqnarray}
m_i \coloneqq \lim_{\delta \to 0^+} \frac{E_F(B_{\nu=i}+2\delta)-E_F(B_{\nu=i}+\delta)\}}{\delta},
\end{eqnarray}
 decreases with higher values of $d_\mathrm{max}$. Specifically, 
 \begin{eqnarray}
 &&m_1= \frac{\hbar e}{2m}\sqrt{1-d_\mathrm{max}^2},\\ 
 &&m_2= \frac{3\hbar e}{2m}\sqrt{1-d_\mathrm{max}^2}.
 \end{eqnarray}
More generally, $m_i = E_{i-1}^{LL}/B$.
The decrease in the slopes $m_1$ and $m_2$ with increasing $d_\mathrm{max}$ causes $B_1$, where the $E_F$ transitions from $E^{LL}_{1}$ to $E^{LL}_{0}$, to increase. When $d_\mathrm{max}$ reaches one, these slopes approach zero, making $B_1$ infinite. This implies that the first Landau level $E^{LL}_{1}$ is always occupied when $E_F^{(0)}>0$ where $E_F^{(0)}$ is the Fermi level at zero magnetic field.

To consider Hall plateaus, we assume the disorder induced Landau level broadening as schematically illustrated in Figure~\ref{fig3}(e) where delocalized electrons exist in the middle of each level while the rest of the states are localized.
{The Hall resistivity $\rho_{xy}$ is quantized as  $\rho_{xy}=-h/(ne^2)$ with a natural number $n$. More explicitly, $n$ is determined by $n=N+1 - \delta_{d_\mathrm{max},1}$, where $N$ is the largest integer that satisfies $E^{(0)}>E^{LL}_N$.}
Figure~$\ref{fig3}$(c) shows the magnetic field dependence of Hall resistivity for $d_\mathrm{max}=0, 0.5, 0.8$ and 1. 
Similar to the oscillating $E_F$, the magnetic field at which $\rho_{xy}$ jumps strongly depends on $d_\mathrm{max}$.
Increasing $d_\mathrm{max}$ extends the length of $n=1$ plateau and shifts the magnetic field $B^*$ where the transition from $n=1$ plateau to $n=0$ plateau occurs. Figure~\ref{fig3}(d) shows the $d_\mathrm{max}$ dependence of $B^*$. When $d_\mathrm{max}=1$, $B^*$ becomes infinity, indicating that the $n=0$ plateau does not exist.

Furthermore, {as shown in the dashed vertical lines} in Fig.~\ref{fig3}(c), one can observe that the jump between $n=1$ and $n=2$ plateaus occurs at higher magnetic fields as $d_\mathrm{max}$ increases when $0\leq d_\mathrm{max}<1$. 
However, when $d_\mathrm{max}=1$, this jump suddenly occurs at a relatively lower magnetic field. This phenomenon originates from the chiral symmetry and the degeneracy between the zeroth and first Landau levels in Eq.~(\ref{eq:LLs}).

More explicitly, the Hall resistivity changes when the Fermi energy passes the delocalized state, as shown in Fig.~\ref{fig3}(e). 
For $d_\mathrm{max}<1$, the transition between the $n=1$ and $n=2$ plateaus occurs when $E_F^{(0)} = E_1^{LL}$. Since $E_1^{LL}(d_\mathrm{max})=\frac{3\hbar e}{2m}B\sqrt{1-d_\mathrm{max}^2}$, the jump requires a higher $B$ as $d_\mathrm{max}$ increases. 
On the other hand, at $d_\mathrm{max}=1$, the transition between the $n=1$ and $n=2$ plateaus happens when $E_F^{(0)} = E_2^{LL}=\frac{\hbar e}{m}\sqrt{2}B$. This is because filling $E_0^{LL}$ and $E_1^{LL}$ corresponds to the jump between $n=-1$ and $n=+1$ plateaus due to chiral symmetry. Note that the system is electrically neutral when both $E^{LL}_0$ and $E^{LL}_1$ are half-filled. Therefore, the transition between $n=1$ and $n=2$ plateaus occurs when $E_F^{(0)} = E_2^{LL}(d_\mathrm{max}=1)$, while for $d_\mathrm{max}<1$ it occurs when $E_F^{(0)} = E_1^{LL}$ [Fig.~\ref{fig3}(e)]. Consequently, the jump between $n=1$ and $n=2$ plateaus for $d_\mathrm{max}=1$ does not follow the trend observed when $d_\mathrm{max}<1$. 
%

\section{Magnetic response functions \label{sec:MRF}}
Magnetic response functions also exhibit strong dependence on $d_\mathrm{max}$. 
We explore how $d_\mathrm{max}$ influences magnetic response functions by utilizing the Roth-Gaou-Niu relation \cite{gao2017zero,roth1966semiclassical,fuchs2018landau} given by
\begin{eqnarray}
(N-\frac{1}{2})\frac{eB}{h}=N_0(\epsilon_F)+BM'_0(\epsilon_F)+\frac{B^2}{2}\chi'_0(\epsilon_F)+O(B^3),\nonumber \\ \label{eq:RGN}
\end{eqnarray}
where $N_0(\epsilon_F)$ is the zero-field integrated density of states at the Fermi energy $\epsilon_F$ (i.e., $N'_0=\partial_\epsilon N_0$ is the density of states),
 $M_0(\epsilon_F)$ is the spontaneous magnetization, and $\chi_0(\epsilon_F)$ is the magnetic susceptibility. Here, the prime denotes the derivative with respect to the energy: $M'_0 =\partial_\epsilon M_0$, $\chi'_0=\partial_\epsilon \chi_0(\epsilon)$. These quantities, which depend on the Fermi energy $\epsilon_F$, are evaluated at zero temperature and in the limit of zero magnetic field. Below, we adopt units where $\hbar=1$ and the flux quantum $\phi_0=h/e=1$.

By applying the results in Eq.~(\ref{eq:LLs}) to Eq.~(\ref{eq:RGN}), we obtain the following expressions {\footnote{Solve $E_N^{LL}=\epsilon_F$ for N, then we get the results.}}
\begin{eqnarray}
&&N_0(\epsilon_F)=\frac{\epsilon_F}{2\pi},\\
&&M’_0(\epsilon_F) = \frac{\Phi_B(d_\mathrm{max})}{2\pi},\\
&&\chi’_0(\epsilon_F) = d_\mathrm{max}^2\frac{\pi}{2\epsilon_F}. \label{eq:Deriv.chi}
\end{eqnarray}
We note that the integrated density of states $N_0(\epsilon_F)$  solely depends on $\epsilon_F$ independent of $d_\mathrm{max}$ because $d_\mathrm{max}$ does not contribute to the energy dispersion. 

The fact that the derivative of $M_0$ is proportional to the Berry phase indicates that the average of the orbital magnetic moment over the Fermi surface vanishes, i.e., $\braket{\mathcal{M}}_{\epsilon_F}=0$. 
This is because, according to the modern theory of magnetization \cite{xiao2010berry,fuchs2018landau,thonhauser2011theory}, differentiating the magnetization with respect to the chemical potential gives
\begin{eqnarray}
M’_0(\epsilon_F)=\braket{\mathcal{M}}_{\epsilon_F} N'_0(\epsilon_F)+\frac{\Phi_B(d_\mathrm{max})}{2\pi},
\end{eqnarray}
where $\braket{\mathcal{M}}_{\epsilon_F}$ represents the average of the orbital magnetic moment over the Fermi surface.
Indeed, for a two-band model with electron-hole symmetry, the orbital magnetic moment is directly related to the Berry curvature: $\mathcal{M}(\bm{k}) =\frac{e}{\hbar} \epsilon_+ (\bm{k})\Omega(\bm{k})$, where $\Omega$ is the Berry curvature \cite{xiao2007valley,fuchs2010topological}.
In a quadratic band touching semimetal, the Berry curvature at finite $\bm{k}$ is always zero while the Berry phase $\Phi_B(d_\mathrm{max})$ can be finite \cite{hwang2021wave}.
Therefore, the average of the orbital magnetic moment over the Fermi surface vanishes, and $M'_0(\epsilon_F)=\frac{\Phi_B(d_\mathrm{max})}{2\pi}$. 

The result for the derivative of the susceptibility agrees with the known results for free electron model and bilayer graphene when $d_\mathrm{max}=0 $ and $d_\mathrm{max}=1$, respectively \cite{fuchs2018landau,safran1984stage,koshino2007orbital}. Considering the form of the susceptibility in the free electron model and bilayer graphene as well as Eq.~(\ref{eq:Deriv.chi})
\footnote{The susceptibility is given $\chi_0(\epsilon_F)=\frac{\pi}{6}$ for free electrons and $\chi_0(\epsilon_F)=\frac{\pi}{2}(\frac{1}{3}+\ln{\frac{|\epsilon_F|}{t}})$ for bilayer graphene\cite{fuchs2018landau,safran1984stage,koshino2007orbital}. The derivative of the susceptibility is $\chi’_0(\epsilon_F) = d_\mathrm{max}^2\frac{\pi}{2\epsilon_F}$ from Eq.~(\ref{eq:Deriv.chi}).}, we propose the following expression for the susceptibility:
\begin{eqnarray}
\chi_0(\epsilon_F)=\frac{\pi}{2}(\frac{1}{3}+d_\mathrm{max}^2\ln{\frac{|\epsilon_F|}{t}}),
\end{eqnarray}
where $t$ is a constant. For bilayer graphene, the constant $t$ is related to interlayer hopping \cite{koshino2007orbital,safran1984stage}. 
One can verify that the derivative of the above equation gives Eq.~(\ref{eq:Deriv.chi}).

\begin{figure}[t]
\includegraphics[width=85mm]{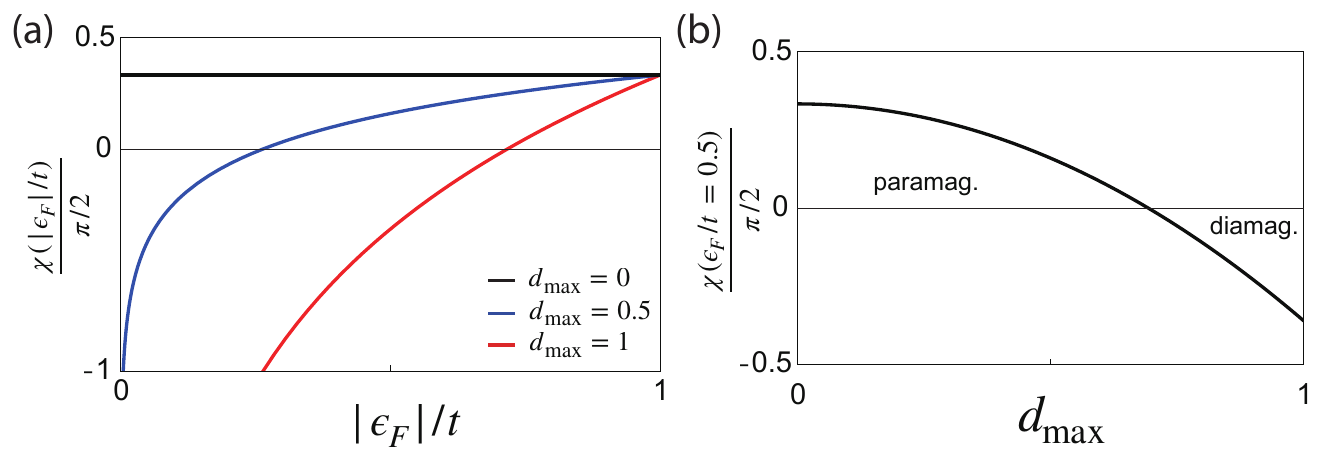} 
\caption{\label{fig5}
(a) $|\epsilon_F|/t$-dependence of the susceptibility $\chi_0$ for $d_\mathrm{max}=0, 0.5$ and 1. (b) $d_\mathrm{max}$-dependence of the susceptibility for $\epsilon_F/t =0.5$.}
\end{figure}

{Furthermore, it can be observed that the sign of $\chi$ changes when $|\epsilon_F| < t$, as shown in Fig.~\ref{fig5}, at the critical value}
\begin{align}
d_\mathrm{max,c}^2 = -\frac{1}{3 \ln (|\epsilon_F|/t)}.
\end{align}
{When $d_\mathrm{max} > d_\mathrm{max,c}$, the system exhibits paramagnetism, i.e., $\chi_0 > 0$. On the other hand, when $0 < d_\mathrm{max} < d_\mathrm{max,c}$, the system exhibits diamagnetism, i.e., $\chi_0 < 0$.
The sign change in $\chi_0$ as $d_\mathrm{max}$ varies indicates that a change in geometry alone, without altering the band structure or Fermi level, can induce a magnetic phase transition.}

\section{Conclusion \label{sec:conclusion}}
To summarize, we have studied the influence of $d_\mathrm{max}$ on Landau levels, the QHE, and magnetic response functions of isotropic quadratic band-touching systems. 
Despite sharing the same energy dispersion, distinct wave function geometry, characterized by different $d_\mathrm{max}$, gives rise to markedly different Landau levels, QHE, and magnetic response functions. 
{To experimentally observe how these physical properties evolve with changes in $d_\mathrm{max}$, it is crucial to identify systems where tuning $d_\mathrm{max}$ is feasible. 
Achieving a non-integer $d_\mathrm{max}$ in momentum space requires $h_x (\bm{k}), h_y (\bm{k}), h_z (\bm{k}) \neq 0$ in the Hamiltonian in the Eq.~(\ref{eq:Ham}) simultaneously. This indicates imaginary and long-range (at least the next nearest neighboring ones) hoppings are necessary in real space,
 suggesting that materials with strong spin-orbit coupling could be potential candidates for $0< d_\mathrm{max} <1$.}
Our finding underscores the significance of quantum geometry of wavefunctions on the magnetic properties of electronic systems. 
Considering the growing interest in the physical responses induced by the quantum geometry, revealing the relation between physical responses and geometry in more general multi-band semimetal systems is an important direction for future study.                

\begin{acknowledgements}
We thank H. Watanabe, M. Koshino, T. Soejima, and J. Jung for the useful discussions.
C-g.O. was supported by Q-STEP, WINGS Program, the University of Tokyo.
J.W.R. was supported by the National Research Foundation of Korea (NRF) Grant
funded by the Korean government (MSIT) (Grant no.
2021R1A2C1010572 and 2021R1A5A1032996 and
2022M3H3A106307411) and the Ministry of Education(Grant no. RS-2023-00285390). B.-J.Y. was supported by
Samsung Science and Technology Foundation under Project No. SSTF-BA2002-06, National Research
Foundation of Korea (NRF) grants funded by the government of Korea (MSIT) (Grants No. NRF-2021R1A5A1032996, and GRDC(Global Research Development Center) Cooperative Hub Program through the National Research Foundation of Korea(NRF) funded by the Ministry of Science and ICT(MSIT) (RS-2023-00258359)").
\end{acknowledgements}

\bibliography{ref.bib}

\begin{thebibliography}{42}%
\makeatletter
\providecommand \@ifxundefined [1]{%
 \@ifx{#1\undefined}
}%
\providecommand \@ifnum [1]{%
 \ifnum #1\expandafter \@firstoftwo
 \else \expandafter \@secondoftwo
 \fi
}%
\providecommand \@ifx [1]{%
 \ifx #1\expandafter \@firstoftwo
 \else \expandafter \@secondoftwo
 \fi
}%
\providecommand \natexlab [1]{#1}%
\providecommand \enquote  [1]{``#1''}%
\providecommand \bibnamefont  [1]{#1}%
\providecommand \bibfnamefont [1]{#1}%
\providecommand \citenamefont [1]{#1}%
\providecommand \href@noop [0]{\@secondoftwo}%
\providecommand \href [0]{\begingroup \@sanitize@url \@href}%
\providecommand \@href[1]{\@@startlink{#1}\@@href}%
\providecommand \@@href[1]{\endgroup#1\@@endlink}%
\providecommand \@sanitize@url [0]{\catcode `\\12\catcode `\$12\catcode
  `\&12\catcode `\#12\catcode `\^12\catcode `\_12\catcode `\%12\relax}%
\providecommand \@@startlink[1]{}%
\providecommand \@@endlink[0]{}%
\providecommand \url  [0]{\begingroup\@sanitize@url \@url }%
\providecommand \@url [1]{\endgroup\@href {#1}{\urlprefix }}%
\providecommand \urlprefix  [0]{URL }%
\providecommand \Eprint [0]{\href }%
\providecommand \doibase [0]{https://doi.org/}%
\providecommand \selectlanguage [0]{\@gobble}%
\providecommand \bibinfo  [0]{\@secondoftwo}%
\providecommand \bibfield  [0]{\@secondoftwo}%
\providecommand \translation [1]{[#1]}%
\providecommand \BibitemOpen [0]{}%
\providecommand \bibitemStop [0]{}%
\providecommand \bibitemNoStop [0]{.\EOS\space}%
\providecommand \EOS [0]{\spacefactor3000\relax}%
\providecommand \BibitemShut  [1]{\csname bibitem#1\endcsname}%
\let\auto@bib@innerbib\@empty
\bibitem [{\citenamefont {Ando}\ and\ \citenamefont
  {Nakanishi}(1998)}]{ando1998impurity}%
  \BibitemOpen
  \bibfield  {author} {\bibinfo {author} {\bibfnamefont {T.}~\bibnamefont
  {Ando}}\ and\ \bibinfo {author} {\bibfnamefont {T.}~\bibnamefont
  {Nakanishi}},\ }\bibfield  {title} {\bibinfo {title} {Impurity scattering in
  carbon nanotubes--absence of back scattering--},\ }\href@noop {} {\bibfield
  {journal} {\bibinfo  {journal} {Journal of the Physical Society of Japan}\
  }\textbf {\bibinfo {volume} {67}},\ \bibinfo {pages} {1704} (\bibinfo {year}
  {1998})}\BibitemShut {NoStop}%
\bibitem [{\citenamefont {Ando}(2005)}]{ando2005theory}%
  \BibitemOpen
  \bibfield  {author} {\bibinfo {author} {\bibfnamefont {T.}~\bibnamefont
  {Ando}},\ }\bibfield  {title} {\bibinfo {title} {Theory of electronic states
  and transport in carbon nanotubes},\ }\href@noop {} {\bibfield  {journal}
  {\bibinfo  {journal} {Journal of the Physical Society of Japan}\ }\textbf
  {\bibinfo {volume} {74}},\ \bibinfo {pages} {777} (\bibinfo {year}
  {2005})}\BibitemShut {NoStop}%
\bibitem [{\citenamefont {Katsnelson}\ \emph {et~al.}(2006)\citenamefont
  {Katsnelson}, \citenamefont {Novoselov},\ and\ \citenamefont
  {Geim}}]{katsnelson2006chiral}%
  \BibitemOpen
  \bibfield  {author} {\bibinfo {author} {\bibfnamefont {M.~I.}\ \bibnamefont
  {Katsnelson}}, \bibinfo {author} {\bibfnamefont {K.~S.}\ \bibnamefont
  {Novoselov}},\ and\ \bibinfo {author} {\bibfnamefont {A.~K.}\ \bibnamefont
  {Geim}},\ }\bibfield  {title} {\bibinfo {title} {Chiral tunnelling and the
  klein paradox in graphene},\ }\href@noop {} {\bibfield  {journal} {\bibinfo
  {journal} {Nature physics}\ }\textbf {\bibinfo {volume} {2}},\ \bibinfo
  {pages} {620} (\bibinfo {year} {2006})}\BibitemShut {NoStop}%
\bibitem [{\citenamefont {Young}\ and\ \citenamefont
  {Kim}(2009)}]{young2009quantum}%
  \BibitemOpen
  \bibfield  {author} {\bibinfo {author} {\bibfnamefont {A.~F.}\ \bibnamefont
  {Young}}\ and\ \bibinfo {author} {\bibfnamefont {P.}~\bibnamefont {Kim}},\
  }\bibfield  {title} {\bibinfo {title} {Quantum interference and klein
  tunnelling in graphene heterojunctions},\ }\href@noop {} {\bibfield
  {journal} {\bibinfo  {journal} {Nature Physics}\ }\textbf {\bibinfo {volume}
  {5}},\ \bibinfo {pages} {222} (\bibinfo {year} {2009})}\BibitemShut {NoStop}%
\bibitem [{\citenamefont {McCann}\ \emph {et~al.}(2006)\citenamefont {McCann},
  \citenamefont {Kechedzhi}, \citenamefont {Fal’ko}, \citenamefont {Suzuura},
  \citenamefont {Ando},\ and\ \citenamefont {Altshuler}}]{mccann2006weak}%
  \BibitemOpen
  \bibfield  {author} {\bibinfo {author} {\bibfnamefont {E.}~\bibnamefont
  {McCann}}, \bibinfo {author} {\bibfnamefont {K.}~\bibnamefont {Kechedzhi}},
  \bibinfo {author} {\bibfnamefont {V.~I.}\ \bibnamefont {Fal’ko}}, \bibinfo
  {author} {\bibfnamefont {H.}~\bibnamefont {Suzuura}}, \bibinfo {author}
  {\bibfnamefont {T.}~\bibnamefont {Ando}},\ and\ \bibinfo {author}
  {\bibfnamefont {B.}~\bibnamefont {Altshuler}},\ }\bibfield  {title} {\bibinfo
  {title} {Weak-localization magnetoresistance and valley symmetry in
  graphene},\ }\href@noop {} {\bibfield  {journal} {\bibinfo  {journal}
  {Physical review letters}\ }\textbf {\bibinfo {volume} {97}},\ \bibinfo
  {pages} {146805} (\bibinfo {year} {2006})}\BibitemShut {NoStop}%
\bibitem [{\citenamefont {Tikhonenko}\ \emph {et~al.}(2009)\citenamefont
  {Tikhonenko}, \citenamefont {Kozikov}, \citenamefont {Savchenko},\ and\
  \citenamefont {Gorbachev}}]{tikhonenko2009transition}%
  \BibitemOpen
  \bibfield  {author} {\bibinfo {author} {\bibfnamefont {F.}~\bibnamefont
  {Tikhonenko}}, \bibinfo {author} {\bibfnamefont {A.}~\bibnamefont {Kozikov}},
  \bibinfo {author} {\bibfnamefont {A.}~\bibnamefont {Savchenko}},\ and\
  \bibinfo {author} {\bibfnamefont {R.}~\bibnamefont {Gorbachev}},\ }\bibfield
  {title} {\bibinfo {title} {Transition between electron localization and
  antilocalization in graphene},\ }\href@noop {} {\bibfield  {journal}
  {\bibinfo  {journal} {Physical Review Letters}\ }\textbf {\bibinfo {volume}
  {103}},\ \bibinfo {pages} {226801} (\bibinfo {year} {2009})}\BibitemShut
  {NoStop}%
\bibitem [{\citenamefont {Lu}\ \emph {et~al.}(2019)\citenamefont {Lu},
  \citenamefont {Stepanov}, \citenamefont {Yang}, \citenamefont {Xie},
  \citenamefont {Aamir}, \citenamefont {Das}, \citenamefont {Urgell},
  \citenamefont {Watanabe}, \citenamefont {Taniguchi}, \citenamefont {Zhang}
  \emph {et~al.}}]{lu2019superconductors}%
  \BibitemOpen
  \bibfield  {author} {\bibinfo {author} {\bibfnamefont {X.}~\bibnamefont
  {Lu}}, \bibinfo {author} {\bibfnamefont {P.}~\bibnamefont {Stepanov}},
  \bibinfo {author} {\bibfnamefont {W.}~\bibnamefont {Yang}}, \bibinfo {author}
  {\bibfnamefont {M.}~\bibnamefont {Xie}}, \bibinfo {author} {\bibfnamefont
  {M.~A.}\ \bibnamefont {Aamir}}, \bibinfo {author} {\bibfnamefont
  {I.}~\bibnamefont {Das}}, \bibinfo {author} {\bibfnamefont {C.}~\bibnamefont
  {Urgell}}, \bibinfo {author} {\bibfnamefont {K.}~\bibnamefont {Watanabe}},
  \bibinfo {author} {\bibfnamefont {T.}~\bibnamefont {Taniguchi}}, \bibinfo
  {author} {\bibfnamefont {G.}~\bibnamefont {Zhang}}, \emph {et~al.},\
  }\bibfield  {title} {\bibinfo {title} {Superconductors, orbital magnets and
  correlated states in magic-angle bilayer graphene},\ }\href@noop {}
  {\bibfield  {journal} {\bibinfo  {journal} {Nature}\ }\textbf {\bibinfo
  {volume} {574}},\ \bibinfo {pages} {653} (\bibinfo {year}
  {2019})}\BibitemShut {NoStop}%
\bibitem [{\citenamefont {Liu}\ \emph {et~al.}(2021)\citenamefont {Liu},
  \citenamefont {Khalaf}, \citenamefont {Lee},\ and\ \citenamefont
  {Vishwanath}}]{liu2021nematic}%
  \BibitemOpen
  \bibfield  {author} {\bibinfo {author} {\bibfnamefont {S.}~\bibnamefont
  {Liu}}, \bibinfo {author} {\bibfnamefont {E.}~\bibnamefont {Khalaf}},
  \bibinfo {author} {\bibfnamefont {J.~Y.}\ \bibnamefont {Lee}},\ and\ \bibinfo
  {author} {\bibfnamefont {A.}~\bibnamefont {Vishwanath}},\ }\bibfield  {title}
  {\bibinfo {title} {Nematic topological semimetal and insulator in magic-angle
  bilayer graphene at charge neutrality},\ }\href@noop {} {\bibfield  {journal}
  {\bibinfo  {journal} {Physical Review Research}\ }\textbf {\bibinfo {volume}
  {3}},\ \bibinfo {pages} {013033} (\bibinfo {year} {2021})}\BibitemShut
  {NoStop}%
\bibitem [{\citenamefont {Cao}\ \emph {et~al.}(2018{\natexlab{a}})\citenamefont
  {Cao}, \citenamefont {Fatemi}, \citenamefont {Fang}, \citenamefont
  {Watanabe}, \citenamefont {Taniguchi}, \citenamefont {Kaxiras},\ and\
  \citenamefont {Jarillo-Herrero}}]{cao2018unconventional}%
  \BibitemOpen
  \bibfield  {author} {\bibinfo {author} {\bibfnamefont {Y.}~\bibnamefont
  {Cao}}, \bibinfo {author} {\bibfnamefont {V.}~\bibnamefont {Fatemi}},
  \bibinfo {author} {\bibfnamefont {S.}~\bibnamefont {Fang}}, \bibinfo {author}
  {\bibfnamefont {K.}~\bibnamefont {Watanabe}}, \bibinfo {author}
  {\bibfnamefont {T.}~\bibnamefont {Taniguchi}}, \bibinfo {author}
  {\bibfnamefont {E.}~\bibnamefont {Kaxiras}},\ and\ \bibinfo {author}
  {\bibfnamefont {P.}~\bibnamefont {Jarillo-Herrero}},\ }\bibfield  {title}
  {\bibinfo {title} {Unconventional superconductivity in magic-angle graphene
  superlattices},\ }\href@noop {} {\bibfield  {journal} {\bibinfo  {journal}
  {Nature}\ }\textbf {\bibinfo {volume} {556}},\ \bibinfo {pages} {43}
  (\bibinfo {year} {2018}{\natexlab{a}})}\BibitemShut {NoStop}%
\bibitem [{\citenamefont {Cao}\ \emph {et~al.}(2018{\natexlab{b}})\citenamefont
  {Cao}, \citenamefont {Fatemi}, \citenamefont {Demir}, \citenamefont {Fang},
  \citenamefont {Tomarken}, \citenamefont {Luo}, \citenamefont
  {Sanchez-Yamagishi}, \citenamefont {Watanabe}, \citenamefont {Taniguchi},
  \citenamefont {Kaxiras} \emph {et~al.}}]{cao2018correlated}%
  \BibitemOpen
  \bibfield  {author} {\bibinfo {author} {\bibfnamefont {Y.}~\bibnamefont
  {Cao}}, \bibinfo {author} {\bibfnamefont {V.}~\bibnamefont {Fatemi}},
  \bibinfo {author} {\bibfnamefont {A.}~\bibnamefont {Demir}}, \bibinfo
  {author} {\bibfnamefont {S.}~\bibnamefont {Fang}}, \bibinfo {author}
  {\bibfnamefont {S.~L.}\ \bibnamefont {Tomarken}}, \bibinfo {author}
  {\bibfnamefont {J.~Y.}\ \bibnamefont {Luo}}, \bibinfo {author} {\bibfnamefont
  {J.~D.}\ \bibnamefont {Sanchez-Yamagishi}}, \bibinfo {author} {\bibfnamefont
  {K.}~\bibnamefont {Watanabe}}, \bibinfo {author} {\bibfnamefont
  {T.}~\bibnamefont {Taniguchi}}, \bibinfo {author} {\bibfnamefont
  {E.}~\bibnamefont {Kaxiras}}, \emph {et~al.},\ }\bibfield  {title} {\bibinfo
  {title} {Correlated insulator behaviour at half-filling in magic-angle
  graphene superlattices},\ }\href@noop {} {\bibfield  {journal} {\bibinfo
  {journal} {Nature}\ }\textbf {\bibinfo {volume} {556}},\ \bibinfo {pages}
  {80} (\bibinfo {year} {2018}{\natexlab{b}})}\BibitemShut {NoStop}%
\bibitem [{\citenamefont {Choi}\ \emph {et~al.}(2021)\citenamefont {Choi},
  \citenamefont {Kim}, \citenamefont {Peng}, \citenamefont {Thomson},
  \citenamefont {Lewandowski}, \citenamefont {Polski}, \citenamefont {Zhang},
  \citenamefont {Arora}, \citenamefont {Watanabe}, \citenamefont {Taniguchi}
  \emph {et~al.}}]{choi2021correlation}%
  \BibitemOpen
  \bibfield  {author} {\bibinfo {author} {\bibfnamefont {Y.}~\bibnamefont
  {Choi}}, \bibinfo {author} {\bibfnamefont {H.}~\bibnamefont {Kim}}, \bibinfo
  {author} {\bibfnamefont {Y.}~\bibnamefont {Peng}}, \bibinfo {author}
  {\bibfnamefont {A.}~\bibnamefont {Thomson}}, \bibinfo {author} {\bibfnamefont
  {C.}~\bibnamefont {Lewandowski}}, \bibinfo {author} {\bibfnamefont
  {R.}~\bibnamefont {Polski}}, \bibinfo {author} {\bibfnamefont
  {Y.}~\bibnamefont {Zhang}}, \bibinfo {author} {\bibfnamefont {H.~S.}\
  \bibnamefont {Arora}}, \bibinfo {author} {\bibfnamefont {K.}~\bibnamefont
  {Watanabe}}, \bibinfo {author} {\bibfnamefont {T.}~\bibnamefont {Taniguchi}},
  \emph {et~al.},\ }\bibfield  {title} {\bibinfo {title} {Correlation-driven
  topological phases in magic-angle twisted bilayer graphene},\ }\href@noop {}
  {\bibfield  {journal} {\bibinfo  {journal} {Nature}\ }\textbf {\bibinfo
  {volume} {589}},\ \bibinfo {pages} {536} (\bibinfo {year}
  {2021})}\BibitemShut {NoStop}%
\bibitem [{\citenamefont {Guerci}\ \emph {et~al.}(2021)\citenamefont {Guerci},
  \citenamefont {Simon},\ and\ \citenamefont {Mora}}]{guerci2021moire}%
  \BibitemOpen
  \bibfield  {author} {\bibinfo {author} {\bibfnamefont {D.}~\bibnamefont
  {Guerci}}, \bibinfo {author} {\bibfnamefont {P.}~\bibnamefont {Simon}},\ and\
  \bibinfo {author} {\bibfnamefont {C.}~\bibnamefont {Mora}},\ }\bibfield
  {title} {\bibinfo {title} {Moir{\'e} lattice effects on the orbital magnetic
  response of twisted bilayer graphene and condon instability},\ }\href@noop {}
  {\bibfield  {journal} {\bibinfo  {journal} {Physical Review B}\ }\textbf
  {\bibinfo {volume} {103}},\ \bibinfo {pages} {224436} (\bibinfo {year}
  {2021})}\BibitemShut {NoStop}%
\bibitem [{\citenamefont {Novoselov}\ \emph {et~al.}(2005)\citenamefont
  {Novoselov}, \citenamefont {Geim}, \citenamefont {Morozov}, \citenamefont
  {Jiang}, \citenamefont {Katsnelson}, \citenamefont {Grigorieva},
  \citenamefont {Dubonos}, \citenamefont {Firsov},\ and\ \citenamefont
  {AA}}]{novoselov2005two}%
  \BibitemOpen
  \bibfield  {author} {\bibinfo {author} {\bibfnamefont {K.~S.}\ \bibnamefont
  {Novoselov}}, \bibinfo {author} {\bibfnamefont {A.~K.}\ \bibnamefont {Geim}},
  \bibinfo {author} {\bibfnamefont {S.~V.}\ \bibnamefont {Morozov}}, \bibinfo
  {author} {\bibfnamefont {D.}~\bibnamefont {Jiang}}, \bibinfo {author}
  {\bibfnamefont {M.~I.}\ \bibnamefont {Katsnelson}}, \bibinfo {author}
  {\bibfnamefont {I.~V.}\ \bibnamefont {Grigorieva}}, \bibinfo {author}
  {\bibfnamefont {S.}~\bibnamefont {Dubonos}}, \bibinfo {author} {\bibnamefont
  {Firsov}},\ and\ \bibinfo {author} {\bibnamefont {AA}},\ }\bibfield  {title}
  {\bibinfo {title} {Two-dimensional gas of massless dirac fermions in
  graphene},\ }\href@noop {} {\bibfield  {journal} {\bibinfo  {journal}
  {nature}\ }\textbf {\bibinfo {volume} {438}},\ \bibinfo {pages} {197}
  (\bibinfo {year} {2005})}\BibitemShut {NoStop}%
\bibitem [{\citenamefont {Zhang}\ \emph {et~al.}(2005)\citenamefont {Zhang},
  \citenamefont {Tan}, \citenamefont {Stormer},\ and\ \citenamefont
  {Kim}}]{zhang2005experimental}%
  \BibitemOpen
  \bibfield  {author} {\bibinfo {author} {\bibfnamefont {Y.}~\bibnamefont
  {Zhang}}, \bibinfo {author} {\bibfnamefont {Y.-W.}\ \bibnamefont {Tan}},
  \bibinfo {author} {\bibfnamefont {H.~L.}\ \bibnamefont {Stormer}},\ and\
  \bibinfo {author} {\bibfnamefont {P.}~\bibnamefont {Kim}},\ }\bibfield
  {title} {\bibinfo {title} {Experimental observation of the quantum hall
  effect and berry's phase in graphene},\ }\href@noop {} {\bibfield  {journal}
  {\bibinfo  {journal} {nature}\ }\textbf {\bibinfo {volume} {438}},\ \bibinfo
  {pages} {201} (\bibinfo {year} {2005})}\BibitemShut {NoStop}%
\bibitem [{\citenamefont {Novoselov}\ \emph {et~al.}(2006)\citenamefont
  {Novoselov}, \citenamefont {McCann}, \citenamefont {Morozov}, \citenamefont
  {Fal’ko}, \citenamefont {Katsnelson}, \citenamefont {Zeitler},
  \citenamefont {Jiang}, \citenamefont {Schedin},\ and\ \citenamefont
  {Geim}}]{novoselov2006unconventional}%
  \BibitemOpen
  \bibfield  {author} {\bibinfo {author} {\bibfnamefont {K.~S.}\ \bibnamefont
  {Novoselov}}, \bibinfo {author} {\bibfnamefont {E.}~\bibnamefont {McCann}},
  \bibinfo {author} {\bibfnamefont {S.}~\bibnamefont {Morozov}}, \bibinfo
  {author} {\bibfnamefont {V.~I.}\ \bibnamefont {Fal’ko}}, \bibinfo {author}
  {\bibfnamefont {M.}~\bibnamefont {Katsnelson}}, \bibinfo {author}
  {\bibfnamefont {U.}~\bibnamefont {Zeitler}}, \bibinfo {author} {\bibfnamefont
  {D.}~\bibnamefont {Jiang}}, \bibinfo {author} {\bibfnamefont
  {F.}~\bibnamefont {Schedin}},\ and\ \bibinfo {author} {\bibfnamefont
  {A.}~\bibnamefont {Geim}},\ }\bibfield  {title} {\bibinfo {title}
  {Unconventional quantum hall effect and berry’s phase of 2$\pi$ in bilayer
  graphene},\ }\href@noop {} {\bibfield  {journal} {\bibinfo  {journal} {Nature
  physics}\ }\textbf {\bibinfo {volume} {2}},\ \bibinfo {pages} {177} (\bibinfo
  {year} {2006})}\BibitemShut {NoStop}%
\bibitem [{\citenamefont {Fal'Ko}(2008)}]{fal2008electronic}%
  \BibitemOpen
  \bibfield  {author} {\bibinfo {author} {\bibfnamefont {V.~I.}\ \bibnamefont
  {Fal'Ko}},\ }\bibfield  {title} {\bibinfo {title} {Electronic properties and
  the quantum hall effect in bilayer graphene},\ }\href@noop {} {\bibfield
  {journal} {\bibinfo  {journal} {Philosophical Transactions of the Royal
  Society A: Mathematical, Physical and Engineering Sciences}\ }\textbf
  {\bibinfo {volume} {366}},\ \bibinfo {pages} {205} (\bibinfo {year}
  {2008})}\BibitemShut {NoStop}%
\bibitem [{\citenamefont {Girvin}\ and\ \citenamefont
  {Prange}(1987)}]{girvin1987quantum}%
  \BibitemOpen
  \bibfield  {author} {\bibinfo {author} {\bibfnamefont {S.}~\bibnamefont
  {Girvin}}\ and\ \bibinfo {author} {\bibfnamefont {R.}~\bibnamefont
  {Prange}},\ }\bibfield  {title} {\bibinfo {title} {The quantum hall effect},\
  }\href@noop {} {\  (\bibinfo {year} {1987})}\BibitemShut {NoStop}%
\bibitem [{\citenamefont {MacDonald}\ and\ \citenamefont
  {MacDonald}(1989)}]{macdonald1989quantum}%
  \BibitemOpen
  \bibfield  {author} {\bibinfo {author} {\bibfnamefont {A.~H.}\ \bibnamefont
  {MacDonald}}\ and\ \bibinfo {author} {\bibfnamefont {A.~H.}\ \bibnamefont
  {MacDonald}},\ }\href@noop {} {\emph {\bibinfo {title} {Quantum Hall effect:
  a perspective}}}\ (\bibinfo  {publisher} {Springer},\ \bibinfo {year}
  {1989})\BibitemShut {NoStop}%
\bibitem [{\citenamefont {Rhim}\ \emph {et~al.}(2020)\citenamefont {Rhim},
  \citenamefont {Kim},\ and\ \citenamefont {Yang}}]{rhim2020quantum}%
  \BibitemOpen
  \bibfield  {author} {\bibinfo {author} {\bibfnamefont {J.-W.}\ \bibnamefont
  {Rhim}}, \bibinfo {author} {\bibfnamefont {K.}~\bibnamefont {Kim}},\ and\
  \bibinfo {author} {\bibfnamefont {B.-J.}\ \bibnamefont {Yang}},\ }\bibfield
  {title} {\bibinfo {title} {Quantum distance and anomalous landau levels of
  flat bands},\ }\href@noop {} {\bibfield  {journal} {\bibinfo  {journal}
  {Nature}\ }\textbf {\bibinfo {volume} {584}},\ \bibinfo {pages} {59}
  (\bibinfo {year} {2020})}\BibitemShut {NoStop}%
\bibitem [{\citenamefont {Rhim}\ and\ \citenamefont
  {Yang}(2021)}]{rhim2021singular}%
  \BibitemOpen
  \bibfield  {author} {\bibinfo {author} {\bibfnamefont {J.-W.}\ \bibnamefont
  {Rhim}}\ and\ \bibinfo {author} {\bibfnamefont {B.-J.}\ \bibnamefont
  {Yang}},\ }\bibfield  {title} {\bibinfo {title} {Singular flat bands},\
  }\href@noop {} {\bibfield  {journal} {\bibinfo  {journal} {Advances in
  Physics: X}\ }\textbf {\bibinfo {volume} {6}},\ \bibinfo {pages} {1901606}
  (\bibinfo {year} {2021})}\BibitemShut {NoStop}%
\bibitem [{\citenamefont {Oh}\ \emph {et~al.}(2022)\citenamefont {Oh},
  \citenamefont {Cho}, \citenamefont {Park},\ and\ \citenamefont
  {Rhim}}]{oh2022bulk}%
  \BibitemOpen
  \bibfield  {author} {\bibinfo {author} {\bibfnamefont {C.-g.}\ \bibnamefont
  {Oh}}, \bibinfo {author} {\bibfnamefont {D.}~\bibnamefont {Cho}}, \bibinfo
  {author} {\bibfnamefont {S.~Y.}\ \bibnamefont {Park}},\ and\ \bibinfo
  {author} {\bibfnamefont {J.-W.}\ \bibnamefont {Rhim}},\ }\bibfield  {title}
  {\bibinfo {title} {Bulk-interface correspondence from quantum distance in
  flat band systems},\ }\href@noop {} {\bibfield  {journal} {\bibinfo
  {journal} {Communications Physics}\ }\textbf {\bibinfo {volume} {5}},\
  \bibinfo {pages} {320} (\bibinfo {year} {2022})}\BibitemShut {NoStop}%
\bibitem [{\citenamefont {Kim}\ \emph {et~al.}(2023)\citenamefont {Kim},
  \citenamefont {Oh},\ and\ \citenamefont {Rhim}}]{kim2023general}%
  \BibitemOpen
  \bibfield  {author} {\bibinfo {author} {\bibfnamefont {H.}~\bibnamefont
  {Kim}}, \bibinfo {author} {\bibfnamefont {C.-g.}\ \bibnamefont {Oh}},\ and\
  \bibinfo {author} {\bibfnamefont {J.-W.}\ \bibnamefont {Rhim}},\ }\bibfield
  {title} {\bibinfo {title} {General construction scheme for geometrically
  nontrivial flat band models},\ }\href@noop {} {\bibfield  {journal} {\bibinfo
   {journal} {arXiv preprint arXiv:2305.00448}\ } (\bibinfo {year}
  {2023})}\BibitemShut {NoStop}%
\bibitem [{\citenamefont {Hwang}\ \emph {et~al.}(2021)\citenamefont {Hwang},
  \citenamefont {Jung}, \citenamefont {Rhim},\ and\ \citenamefont
  {Yang}}]{hwang2021wave}%
  \BibitemOpen
  \bibfield  {author} {\bibinfo {author} {\bibfnamefont {Y.}~\bibnamefont
  {Hwang}}, \bibinfo {author} {\bibfnamefont {J.}~\bibnamefont {Jung}},
  \bibinfo {author} {\bibfnamefont {J.-W.}\ \bibnamefont {Rhim}},\ and\
  \bibinfo {author} {\bibfnamefont {B.-J.}\ \bibnamefont {Yang}},\ }\bibfield
  {title} {\bibinfo {title} {Wave-function geometry of band crossing points in
  two dimensions},\ }\href@noop {} {\bibfield  {journal} {\bibinfo  {journal}
  {Physical Review B}\ }\textbf {\bibinfo {volume} {103}},\ \bibinfo {pages}
  {L241102} (\bibinfo {year} {2021})}\BibitemShut {NoStop}%
\bibitem [{\citenamefont {Jung}\ \emph {et~al.}(2024)\citenamefont {Jung},
  \citenamefont {Lim},\ and\ \citenamefont {Yang}}]{jung2024quantum}%
  \BibitemOpen
  \bibfield  {author} {\bibinfo {author} {\bibfnamefont {J.}~\bibnamefont
  {Jung}}, \bibinfo {author} {\bibfnamefont {H.}~\bibnamefont {Lim}},\ and\
  \bibinfo {author} {\bibfnamefont {B.-J.}\ \bibnamefont {Yang}},\ }\bibfield
  {title} {\bibinfo {title} {Quantum geometry and landau levels of quadratic
  band crossings},\ }\href@noop {} {\bibfield  {journal} {\bibinfo  {journal}
  {Physical Review B}\ }\textbf {\bibinfo {volume} {109}},\ \bibinfo {pages}
  {035134} (\bibinfo {year} {2024})}\BibitemShut {NoStop}%
\bibitem [{\citenamefont {Gao}\ and\ \citenamefont {Niu}(2017)}]{gao2017zero}%
  \BibitemOpen
  \bibfield  {author} {\bibinfo {author} {\bibfnamefont {Y.}~\bibnamefont
  {Gao}}\ and\ \bibinfo {author} {\bibfnamefont {Q.}~\bibnamefont {Niu}},\
  }\bibfield  {title} {\bibinfo {title} {Zero-field magnetic response functions
  in landau levels},\ }\href@noop {} {\bibfield  {journal} {\bibinfo  {journal}
  {Proceedings of the National Academy of Sciences}\ }\textbf {\bibinfo
  {volume} {114}},\ \bibinfo {pages} {7295} (\bibinfo {year}
  {2017})}\BibitemShut {NoStop}%
\bibitem [{\citenamefont {Roth}(1966)}]{roth1966semiclassical}%
  \BibitemOpen
  \bibfield  {author} {\bibinfo {author} {\bibfnamefont {L.~M.}\ \bibnamefont
  {Roth}},\ }\bibfield  {title} {\bibinfo {title} {Semiclassical theory of
  magnetic energy levels and magnetic susceptibility of bloch electrons},\
  }\href@noop {} {\bibfield  {journal} {\bibinfo  {journal} {Physical Review}\
  }\textbf {\bibinfo {volume} {145}},\ \bibinfo {pages} {434} (\bibinfo {year}
  {1966})}\BibitemShut {NoStop}%
\bibitem [{\citenamefont {Fuchs}\ \emph {et~al.}(2018)\citenamefont {Fuchs},
  \citenamefont {Pi{\'e}chon},\ and\ \citenamefont
  {Montambaux}}]{fuchs2018landau}%
  \BibitemOpen
  \bibfield  {author} {\bibinfo {author} {\bibfnamefont {J.-N.}\ \bibnamefont
  {Fuchs}}, \bibinfo {author} {\bibfnamefont {F.}~\bibnamefont {Pi{\'e}chon}},\
  and\ \bibinfo {author} {\bibfnamefont {G.}~\bibnamefont {Montambaux}},\
  }\bibfield  {title} {\bibinfo {title} {Landau levels, response functions and
  magnetic oscillations from a generalized onsager relation},\ }\href@noop {}
  {\bibfield  {journal} {\bibinfo  {journal} {SciPost Physics}\ }\textbf
  {\bibinfo {volume} {4}},\ \bibinfo {pages} {024} (\bibinfo {year}
  {2018})}\BibitemShut {NoStop}%
\bibitem [{\citenamefont {Koshino}\ and\ \citenamefont
  {Ando}(2007)}]{koshino2007orbital}%
  \BibitemOpen
  \bibfield  {author} {\bibinfo {author} {\bibfnamefont {M.}~\bibnamefont
  {Koshino}}\ and\ \bibinfo {author} {\bibfnamefont {T.}~\bibnamefont {Ando}},\
  }\bibfield  {title} {\bibinfo {title} {Orbital diamagnetism in multilayer
  graphenes: Systematic study with the effective mass approximation},\
  }\href@noop {} {\bibfield  {journal} {\bibinfo  {journal} {Physical Review
  B}\ }\textbf {\bibinfo {volume} {76}},\ \bibinfo {pages} {085425} (\bibinfo
  {year} {2007})}\BibitemShut {NoStop}%
\bibitem [{\citenamefont {McCann}\ and\ \citenamefont
  {Koshino}(2013)}]{mccann2013electronic}%
  \BibitemOpen
  \bibfield  {author} {\bibinfo {author} {\bibfnamefont {E.}~\bibnamefont
  {McCann}}\ and\ \bibinfo {author} {\bibfnamefont {M.}~\bibnamefont
  {Koshino}},\ }\bibfield  {title} {\bibinfo {title} {The electronic properties
  of bilayer graphene},\ }\href@noop {} {\bibfield  {journal} {\bibinfo
  {journal} {Reports on Progress in physics}\ }\textbf {\bibinfo {volume}
  {76}},\ \bibinfo {pages} {056503} (\bibinfo {year} {2013})}\BibitemShut
  {NoStop}%
\bibitem [{\citenamefont {Provost}\ and\ \citenamefont
  {Vallee}(1980)}]{provost1980riemannian}%
  \BibitemOpen
  \bibfield  {author} {\bibinfo {author} {\bibfnamefont {J.}~\bibnamefont
  {Provost}}\ and\ \bibinfo {author} {\bibfnamefont {G.}~\bibnamefont
  {Vallee}},\ }\bibfield  {title} {\bibinfo {title} {Riemannian structure on
  manifolds of quantum states},\ }\href@noop {} {\bibfield  {journal} {\bibinfo
   {journal} {Communications in Mathematical Physics}\ }\textbf {\bibinfo
  {volume} {76}},\ \bibinfo {pages} {289} (\bibinfo {year} {1980})}\BibitemShut
  {NoStop}%
\bibitem [{\citenamefont {McCann}\ and\ \citenamefont
  {Fal’ko}(2006)}]{mccann2006landau}%
  \BibitemOpen
  \bibfield  {author} {\bibinfo {author} {\bibfnamefont {E.}~\bibnamefont
  {McCann}}\ and\ \bibinfo {author} {\bibfnamefont {V.~I.}\ \bibnamefont
  {Fal’ko}},\ }\bibfield  {title} {\bibinfo {title} {Landau-level degeneracy
  and quantum hall effect in a graphite bilayer},\ }\href@noop {} {\bibfield
  {journal} {\bibinfo  {journal} {Physical review letters}\ }\textbf {\bibinfo
  {volume} {96}},\ \bibinfo {pages} {086805} (\bibinfo {year}
  {2006})}\BibitemShut {NoStop}%
\bibitem [{\citenamefont {Landau}(1930)}]{landau1930diamagnetismus}%
  \BibitemOpen
  \bibfield  {author} {\bibinfo {author} {\bibfnamefont {L.}~\bibnamefont
  {Landau}},\ }\bibfield  {title} {\bibinfo {title} {Diamagnetismus der
  metalle},\ }\href@noop {} {\bibfield  {journal} {\bibinfo  {journal}
  {Zeitschrift f{\"u}r Physik}\ }\textbf {\bibinfo {volume} {64}},\ \bibinfo
  {pages} {629} (\bibinfo {year} {1930})}\BibitemShut {NoStop}%
\bibitem [{\citenamefont {Koshino}\ and\ \citenamefont
  {Ando}(2010)}]{koshino2010anomalous}%
  \BibitemOpen
  \bibfield  {author} {\bibinfo {author} {\bibfnamefont {M.}~\bibnamefont
  {Koshino}}\ and\ \bibinfo {author} {\bibfnamefont {T.}~\bibnamefont {Ando}},\
  }\bibfield  {title} {\bibinfo {title} {Anomalous orbital magnetism in
  dirac-electron systems: role of pseudospin paramagnetism},\ }\href@noop {}
  {\bibfield  {journal} {\bibinfo  {journal} {Physical Review B}\ }\textbf
  {\bibinfo {volume} {81}},\ \bibinfo {pages} {195431} (\bibinfo {year}
  {2010})}\BibitemShut {NoStop}%
\bibitem [{Note1()}]{Note1}%
  \BibitemOpen
  \bibinfo {note} {Solve $E_N^{LL}=\epsilon _F$ for N, then we get the
  results.}\BibitemShut {Stop}%
\bibitem [{\citenamefont {Xiao}\ \emph {et~al.}(2010)\citenamefont {Xiao},
  \citenamefont {Chang},\ and\ \citenamefont {Niu}}]{xiao2010berry}%
  \BibitemOpen
  \bibfield  {author} {\bibinfo {author} {\bibfnamefont {D.}~\bibnamefont
  {Xiao}}, \bibinfo {author} {\bibfnamefont {M.-C.}\ \bibnamefont {Chang}},\
  and\ \bibinfo {author} {\bibfnamefont {Q.}~\bibnamefont {Niu}},\ }\bibfield
  {title} {\bibinfo {title} {Berry phase effects on electronic properties},\
  }\href@noop {} {\bibfield  {journal} {\bibinfo  {journal} {Reviews of modern
  physics}\ }\textbf {\bibinfo {volume} {82}},\ \bibinfo {pages} {1959}
  (\bibinfo {year} {2010})}\BibitemShut {NoStop}%
\bibitem [{\citenamefont {Thonhauser}(2011)}]{thonhauser2011theory}%
  \BibitemOpen
  \bibfield  {author} {\bibinfo {author} {\bibfnamefont {T.}~\bibnamefont
  {Thonhauser}},\ }\bibfield  {title} {\bibinfo {title} {Theory of orbital
  magnetization in solids},\ }\href@noop {} {\bibfield  {journal} {\bibinfo
  {journal} {International Journal of Modern Physics B}\ }\textbf {\bibinfo
  {volume} {25}},\ \bibinfo {pages} {1429} (\bibinfo {year}
  {2011})}\BibitemShut {NoStop}%
\bibitem [{\citenamefont {Xiao}\ \emph {et~al.}(2007)\citenamefont {Xiao},
  \citenamefont {Yao},\ and\ \citenamefont {Niu}}]{xiao2007valley}%
  \BibitemOpen
  \bibfield  {author} {\bibinfo {author} {\bibfnamefont {D.}~\bibnamefont
  {Xiao}}, \bibinfo {author} {\bibfnamefont {W.}~\bibnamefont {Yao}},\ and\
  \bibinfo {author} {\bibfnamefont {Q.}~\bibnamefont {Niu}},\ }\bibfield
  {title} {\bibinfo {title} {Valley-contrasting physics in graphene: magnetic
  moment and topological transport},\ }\href@noop {} {\bibfield  {journal}
  {\bibinfo  {journal} {Physical review letters}\ }\textbf {\bibinfo {volume}
  {99}},\ \bibinfo {pages} {236809} (\bibinfo {year} {2007})}\BibitemShut
  {NoStop}%
\bibitem [{\citenamefont {Fuchs}\ \emph {et~al.}(2010)\citenamefont {Fuchs},
  \citenamefont {Pi{\'e}chon}, \citenamefont {Goerbig},\ and\ \citenamefont
  {Montambaux}}]{fuchs2010topological}%
  \BibitemOpen
  \bibfield  {author} {\bibinfo {author} {\bibfnamefont {J.}~\bibnamefont
  {Fuchs}}, \bibinfo {author} {\bibfnamefont {F.}~\bibnamefont {Pi{\'e}chon}},
  \bibinfo {author} {\bibfnamefont {M.}~\bibnamefont {Goerbig}},\ and\ \bibinfo
  {author} {\bibfnamefont {G.}~\bibnamefont {Montambaux}},\ }\bibfield  {title}
  {\bibinfo {title} {Topological berry phase and semiclassical quantization of
  cyclotron orbits for two dimensional electrons in coupled band models},\
  }\href@noop {} {\bibfield  {journal} {\bibinfo  {journal} {The European
  Physical Journal B}\ }\textbf {\bibinfo {volume} {77}},\ \bibinfo {pages}
  {351} (\bibinfo {year} {2010})}\BibitemShut {NoStop}%
\bibitem [{\citenamefont {Safran}(1984)}]{safran1984stage}%
  \BibitemOpen
  \bibfield  {author} {\bibinfo {author} {\bibfnamefont {S.}~\bibnamefont
  {Safran}},\ }\bibfield  {title} {\bibinfo {title} {Stage dependence of
  magnetic susceptibility of intercalated graphite},\ }\href@noop {} {\bibfield
   {journal} {\bibinfo  {journal} {Physical Review B}\ }\textbf {\bibinfo
  {volume} {30}},\ \bibinfo {pages} {421} (\bibinfo {year} {1984})}\BibitemShut
  {NoStop}%
\bibitem [{Note2()}]{Note2}%
  \BibitemOpen
  \bibinfo {note} {The susceptibility is given $\chi _0(\epsilon _F)=\protect
  \frac {\pi }{6}$ for free electrons and $\chi _0(\epsilon _F)=\protect \frac
  {\pi }{2}(\protect \frac {1}{3}+\protect \qopname \relax o{ln}{\protect \frac
  {|\epsilon _F|}{t}})$ for bilayer graphene\cite
  {fuchs2018landau,safran1984stage,koshino2007orbital}. The derivative of the
  susceptibility is $\chi ’_0(\epsilon _F) = d_\protect \mathrm
  {max}^2\protect \frac {\pi }{2\epsilon _F}$ from Eq.~(\ref
  {eq:Deriv.chi}).}\BibitemShut {Stop}%
\bibitem [{\citenamefont {Pi{\'e}chon}\ \emph {et~al.}(2016)\citenamefont
  {Pi{\'e}chon}, \citenamefont {Raoux}, \citenamefont {Fuchs},\ and\
  \citenamefont {Montambaux}}]{piechon2016geometric}%
  \BibitemOpen
  \bibfield  {author} {\bibinfo {author} {\bibfnamefont {F.}~\bibnamefont
  {Pi{\'e}chon}}, \bibinfo {author} {\bibfnamefont {A.}~\bibnamefont {Raoux}},
  \bibinfo {author} {\bibfnamefont {J.-N.}\ \bibnamefont {Fuchs}},\ and\
  \bibinfo {author} {\bibfnamefont {G.}~\bibnamefont {Montambaux}},\ }\bibfield
   {title} {\bibinfo {title} {Geometric orbital susceptibility: Quantum metric
  without berry curvature},\ }\href@noop {} {\bibfield  {journal} {\bibinfo
  {journal} {Physical Review B}\ }\textbf {\bibinfo {volume} {94}},\ \bibinfo
  {pages} {134423} (\bibinfo {year} {2016})}\BibitemShut {NoStop}%
\bibitem [{\citenamefont {Ozawa}\ and\ \citenamefont
  {Mera}(2021)}]{ozawa2021relations}%
  \BibitemOpen
  \bibfield  {author} {\bibinfo {author} {\bibfnamefont {T.}~\bibnamefont
  {Ozawa}}\ and\ \bibinfo {author} {\bibfnamefont {B.}~\bibnamefont {Mera}},\
  }\bibfield  {title} {\bibinfo {title} {Relations between topology and the
  quantum metric for chern insulators},\ }\href@noop {} {\bibfield  {journal}
  {\bibinfo  {journal} {Physical Review B}\ }\textbf {\bibinfo {volume}
  {104}},\ \bibinfo {pages} {045103} (\bibinfo {year} {2021})}\BibitemShut
  {NoStop}%
\end{thebibliography}%

\onecolumngrid

\clearpage

\appendix

\section{Landau levels of the continuum model}
The continuum Hamiltonian is given by
    \begin{eqnarray}
        \mathcal{H}_{0}({\bm{k}}) = \sum_{\alpha } h_\alpha ({\bm{k}}) \sigma_\alpha , \label{Seq:Ham}
    \end{eqnarray}
where $\sigma_\alpha$ represents an identity ($\alpha=0$) and Pauli matrices ($\alpha = x,y,z$).
Here, $h_\alpha ({\bm{k}})$ is a real quadratic function: $h_{x} ({\bm{k}}) = {d\sqrt{1-d^2}} k_y^2,~h_{y} ({\bm{k}}) = d k_x k_y,~h_{z} ({\bm{k}}) = k_x^2/2+(1-2d^2)k_y^2/2$, and $h_{0} ({\bm{k}}) = 0$. 
This Hamiltonian is obtained from the continuum Hamiltonian in Eq.~(\ref{eq:Ham}) by a unitary transformation with $U=\frac{1-i \sigma_y}{\sqrt{1}}$.

We analyze the Landau levels of the continuum Hamiltonian with $\xi=1$ after the replacement $k_x \to (a+a^\dagger)/(\sqrt{2}l_B)$ and $k_y \to i(a-a^\dagger)/(\sqrt{2}l_B)$, where $l_B=\sqrt{\hbar/eB}$ is a magnetic length, and $a$, $a^\dagger$ are the annihilation and creation operators, respectively.
To ensure the Hamiltonian's hermiticity, a symmetrization is performed: $k_xk_y=(k_x k_y+k_y k_x)/2=i(a^2-(a^\dagger)^2)/(2l_B^2)$. Then, the continuum Hamiltonian in Eq.~(\ref{Seq:Ham}) is transformed to
\begin{eqnarray}
H_{LL}=\frac{1}{2l_B^2}  \bpm h_{11} & h_{12}\\ h_{21} & h_{22} \epm, \label{Seq:HLL}
\end{eqnarray}
where 
\begin{eqnarray}
 &&h_{11}=d_\mathrm{max}^2 (a^2+a^{\dagger 2})+(1-d_\mathrm{max}^2)(2a^{\dagger}a+1),\\
 &&h_{12}=d_\mathrm{max}(1-\sqrt{1-d_\mathrm{max}^2})a^2-d_\mathrm{max}(1+\sqrt{1-d_\mathrm{max}^2})a^{\dagger 2}+d_\mathrm{max}\sqrt{1-d_\mathrm{max}^2}(2a^{\dagger}a+1)=h_{21}^\dagger,\\
 &&h_{22}= -d_\mathrm{max}^2 (a^2+a^{\dagger 2})-(1-d_\mathrm{max}^2)(2a^\dagger a+1).
\end{eqnarray}
Note that this Hamiltonian has a chiral symmetry $\{-\sigma_x,H_{LL} \}=-H_{LL}$, where $-\sigma_x= U^{-1} \sigma_z U$, when $d_\mathrm{max}=0$ or $d_\mathrm{max}=1$. 

One can solve this problem using the following wavefunction:
\begin{eqnarray}
\ket{\psi}= \sum_{n=0}^{\infty} v_n \ket{u_n}=\sum_{n=0}^{\infty} \bpm C_n\\ D_n\epm \ket{u_n}, \label{seq:wf}
\end{eqnarray}
where $\ket{u_n}$ is a normalized state satisfying $a\ket{u_n}=\sqrt{n}\ket{u_{n-1}}$ and $a^\dagger\ket{u_n}=\sqrt{n+1}\ket{u_{n+1}}$, and $C_n$ and $D_n$ are complex coefficients. Using this wavefunction, the Hamiltonian in Eq.~(\ref{Seq:HLL}) can be described as 
\begin{eqnarray}
H_{LL}=\frac{1}{2l_B^2} = \bpm h_0 &0&g_0&0&...\\
0&h_1&0&g_1&...\\ 
g_0^\dagger&0&h_2 & 0&...\\
0&g_1^\dagger&0&h_3&...\\
:&:&:&:&:\epm,
\end{eqnarray}
where
\begin{eqnarray}
h_n= (2n+1) \bpm 1-d_\mathrm{max}^2 & d_\mathrm{max}\sqrt{1-d_\mathrm{max}^2}\\
d_\mathrm{max}\sqrt{1-d_\mathrm{max}^2}&-(1-d_\mathrm{max}^2)
\epm,
\end{eqnarray}
and
\begin{eqnarray}
g_n= \sqrt{(n+1)(n+2)} \bpm d_\mathrm{max}^2 & d_\mathrm{max}(1-\sqrt{1-d_\mathrm{max}^2})\\
-d_\mathrm{max}(1+\sqrt{1-d_\mathrm{max}^2})&-d_\mathrm{max}^2
\epm.
\end{eqnarray}
From this Hamiltonian, one can get 
\begin{eqnarray}
\frac{1}{2l_B^2}\bpm h_{n} &g_{n}\\
g_{n}^\dagger &h_{n+2} \epm  \bpm v_{n} \\v_{n+2}\epm =E\bpm v_{n}\\v_{n+2}\epm.
\end{eqnarray}
By using $v_{n+2} = [\tilde{E}-h_{n+2}]^{-1}g^\dagger_{n}v_{n}$, where $\tilde{E}=2l_B^2 E$, one can obtain the following equation:
\begin{eqnarray}
\frac{1}{2l_B^2}[h_{n}+g_{n}(\tilde{E}-h_{n+2})^{-1}g^\dagger_{n}]v_{n}=Ev_{n}.
\end{eqnarray}
Since $n\leq 1$ this equation does not hold because $n$-th Landau level only comes from the $n+2$-th Landau levels and not from the $n-2$-th Landau levels. Thus, we consider the zero-th and first order energies seperately. Calculating these, one can get the Landau levels in Eq~(6) in the main text. 

\newpage
\section{Landau levels of a square lattice}
\begin{figure*}[t]
\includegraphics[width=\textwidth]{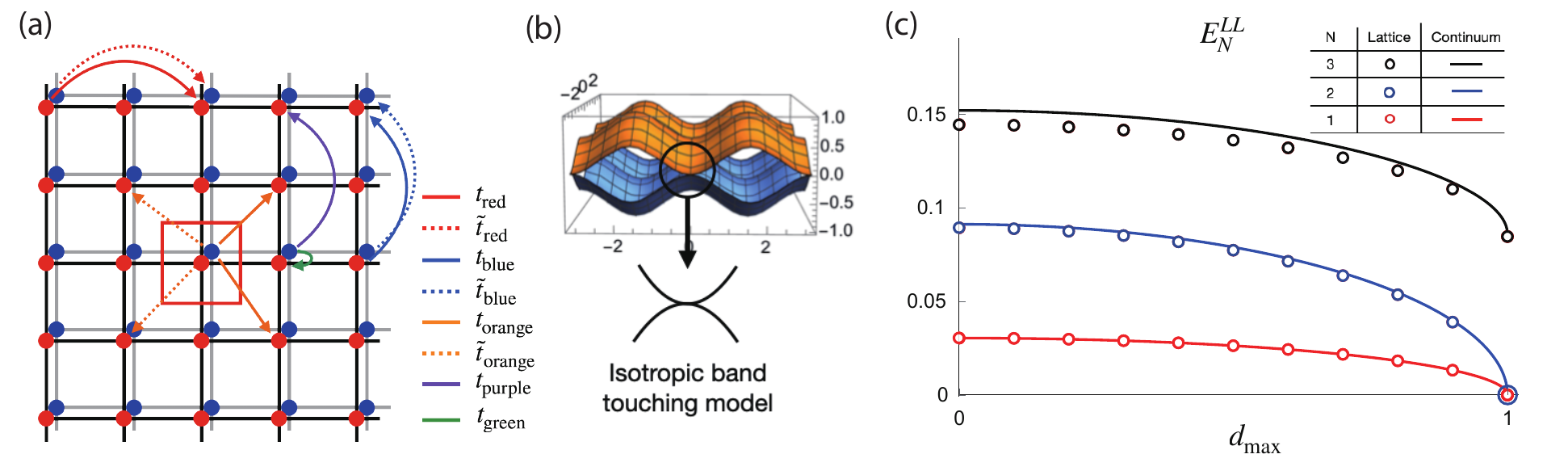} 
\caption{\label{fig4}
(a) Lattice and hopping structure of the square lattice model. Here, $t$'s and $\tilde{t}$'s are the hopping parameters. In this model, $d_\mathrm{max}$ can be varied from 0 to 1 while maintaining the band structure by changing the hopping parameters. (b) Band structure of the lattice model. (c) Zero-th, first and second Landau levels of the lattice model and the continuum model in Eq~(6) in the main text. The circles correspond to the lattice model, and the solid lines are from the continuum model.}
\end{figure*}

To confirm our prediction based on a continuum model, we consider a lattice model on the square lattice whose effective Hamiltonian at the $\Gamma$ point is given as Eq.~(3) in the main text with $\xi=+1$. We consider various long-range hopping processes as illustrated in Fig.~\ref{fig4}(a).
The hopping parameters are given by $t_\mathrm{red} = -\tilde{t}_\mathrm{red}=-1/8$, $t_\mathrm{blue}=-\tilde{t}_{\mathrm{blue}}=-1/8+d_\mathrm{max}$, $t_\mathrm{orange}=-\tilde{t}_\mathrm{orange}=i d_\mathrm{max}/4$, $t_\mathrm{purple}=-d_\mathrm{max}\sqrt{1-d_\mathrm{max}^2}/4$, and $t_\mathrm{green}=d_\mathrm{max}\sqrt{1-d_\mathrm{max}^2}/2$. 
The explicit form of the tight-binding Hamiltonian for this model is 
\begin{eqnarray}
H_\text{Lattice} &=&\sum_{m,n} \frac{1-d_\mathrm{max}^2}{2}(A^\dagger_{m,n} A_{m,n} -B^\dagger_{m,n}B_{m,n})+\Bigg[t_\mathrm{green}A^\dagger_{m,n}B_{m,n}+t_\mathrm{red}(A^\dagger_{m+2,n} A_{m,n} -B^\dagger_{m+2,n}B_{m,n}) \nonumber \\
&&+t_\mathrm{blue}(A^\dagger_{m,n+2} A_{m,n} -B^\dagger_{m,n+2}B_{m,n})+
t_\mathrm{purple}(A^\dagger_{m,n+2}B_{m,n}+A^\dagger_{m,n-2}B_{m,n})
\nonumber \\
&&+t_\mathrm{orange}(A^\dagger_{m+1,n+1}B_{m,n}+A^\dagger_{m-1,n-1}B_{m,n}-A^\dagger_{m-1,n+1}B_{m,n}-A^\dagger_{m+1,n-1}B_{m,n})+h.c.\Bigg].
\end{eqnarray}
The lattice Hamiltonian has the energy eigenvalues $E=\pm(2-\cos{2 k_x}-\cos{2k_y})/4$, shown in Fig.~\ref{fig4}(b), which remain invariant under changes in $d_\mathrm{max}$($0 \leq d_\mathrm{max} \leq 1$).

Our predictions of Landau levels in Eq.~(6) in the main text are confirmed by this lattice model.
We consider commensurate magnetic fluxes $\phi$ satisfying $\phi/\phi_0 = 1/q$, where $q$ is a natural number and $\phi_0$ is the flux quantum. Figure~\ref{fig4}(c) depicts the $d_\mathrm{max}$-dependence of the zero-th, first, and second Landau levels. 
One can verify that our analytic results match well with the results of the lattice model.

\newpage
\section{Landau levels under semiclassical approximation}

{
Physical observables related to Landau levels are closely linked to the geometry of eigenstates and can therefore be directly computed from the wavefunctions. For example, the derivative of the magnetization in Eq.~(16) of the main text is described by the Berry phase, a property of the Bloch wavefunction in the absence of a magnetic field. Similarly, Hall conductivity can be obtained from wavefunctions in the presence of a magnetic field.
In this section, we show that the physical observables derived from the Landau levels using the Roth-Gaou-Niu relation \cite{gao2017zero,roth1966semiclassical,fuchs2018landau}—such as Hall conductivity $\sigma_{xy}$, the derivative of magnetization $M'_0(\epsilon_F)$, and the derivative of susceptibility $\chi'_0(\epsilon_F)$—can be derived from the geometric properties of the wavefunction.}

{The Berry phase, the quantum geometric tensor and the Berry curvature for the model in Eq.~(3) in the main text are given by
\begin{eqnarray}
    &&\Phi_B = -2\pi\xi\sqrt{1-d_\mathrm{max}^2}~~~~~(\mathrm{mod} 2\pi),   \label{eq:BP}\\
    &&g^n_{xx}(\bm{k})=2d_\mathrm{max}^2\frac{k_y^2}{k^4}, ~~g^n_{yy}(\bm{k})=2d_\mathrm{max}^2\frac{k_x^2}{k^4},\nonumber \\
    &&g^n_{xy}(\bm{k})=g^n_{yx}(\bm{k})=-2d_\mathrm{max}^2\frac{k_xk_y}{k^4}, \nonumber \\
    &&\Omega^n_{xy} (\bm{k})=0,
\end{eqnarray}
where $n =\pm$ is band index.
Here, we focus on the case of $\xi=+1$.
}

{
First, let us consider the derivative of magnetization. 
According to the modern theory of magnetization \cite{xiao2010berry,fuchs2018landau,thonhauser2011theory}, differentiating the magnetization with respect to the chemical potential gives
\begin{eqnarray}
M’_0(\epsilon_F)=\braket{\mathcal{M}}_{\epsilon_F} N'_0(\epsilon_F)+\frac{\Phi_B(d_\mathrm{max})}{2\pi},
\end{eqnarray}
where $\braket{\mathcal{M}}_{\epsilon_F}$ represents the average of the orbital magnetic moment over the Fermi surface.
Indeed, for a two-band model with electron-hole symmetry, $\mathcal{M}(\bm{k}) =\frac{e}{\hbar} \epsilon_+ (\bm{k})\Omega(\bm{k})$ \cite{xiao2007valley,fuchs2010topological}.
Therefore, the average of the orbital magnetic moment over the Fermi surface vanishes, and $M'_0(\epsilon_F)={\Phi_B(d_\mathrm{max})}/({2\pi})$. 
}

{
Next, let us focus on the derivative of the susceptibility.
The interband contribution of the susceptibility can be decomposed into three terms \cite{piechon2016geometric}: 
\begin{align}
\chi_{\mathrm{inter}}= \chi_\Omega + \chi_g + \tilde{\chi}_g.
\end{align}
The first term $\chi_\Omega$ is related to the Berry curvature. Since the Berry curvature of the model is zero, this term is negligable.
The third term $\tilde{\chi}_g$ only appears in the absence of electron-hole symmetry, and is also negligible in our system.
The fundamental contribution comes from $\chi_g$, which is related to the quantum metric. The explicit form is given by
\begin{align}
\chi_g (\epsilon_F) = -\sum_{i,j = x,y}\sum_{n=\pm} \sum_{\bm{k}} n \frac{\Theta(\epsilon_F - \epsilon_n(\bm{k}))}{2\epsilon_{+} (\bm{k})}\partial_j \left(\epsilon_+(\bm{k})^2 \partial_i g_{ij}(\bm{k}) \right).
\end{align}
Differnetiating this gives:
\begin{align}
\chi'_g(\epsilon_F) =  -\sum_{i,j = x,y}\sum_{n=\pm} \sum_{\bm{k}} n \frac{\delta(\epsilon_F - \epsilon_n(\bm{k}))}{2\epsilon_{+} (\bm{k})}\partial_j \left(\epsilon_+(\bm{k})^2 \partial_i g_{ij}(\bm{k}) \right).
\end{align}
After straightforward calculation, we find $\chi'_g = d_\mathrm{max}^2\frac{\pi}{2\epsilon_F}$.
}

{
Finally, we calculate the Hall conductivity. To do this, we need the geometric quantitiy of the wavefunction under a magnetic field.
For a given $m$-th Landau level, the wavefunction in Eq.~(\ref{seq:wf}) is written as 
\begin{align}
\ket{\psi_m} = N(m)\left(v_m \ket{u_m}+ v_{m+2}\ket{u_{m+2}}\right),
\end{align}
where $N(m)$ represents normalization factor. 
If we take a unitary transformation, this can be written as
\begin{align}
\ket{\psi_m} =\frac{1}{\sqrt{1+f(m)^2}}\left(\begin{pmatrix} 0\\1 \end{pmatrix} \ket{u_m}+ \begin{pmatrix} f(m)\\0 \end{pmatrix}\ket{u_{m+2}}\right),
\end{align}
where $f(m)=-(3\sqrt{1-d_\mathrm{max}^2}+2\sqrt{1-d_\mathrm{max}^2}+\sqrt{(2n+3)^2-d_\mathrm{max}^2})/(2d_\mathrm{max}\sqrt{(n+1)(n+2)})$.
To calculate the geometric properties of wavefunctions under a magnetic field, we consider magnetic translation symmtery, and the periodic part of the eigenfunction $\ket{\psi_m}$, with a momentum $\bm{k}$, which we denote as $\ket{\bar{\psi}_{m,\bm{k}}}$. We follow  Ref.~\cite{ozawa2021relations}, and calculate the followings:
\begin{align}
&\langle \bar{\psi}_{m,k} | \partial_x \bar{\psi}_{n,k} \rangle = -i \left( \frac{l_B}{2} \delta_{m+1,n} + k_y l_B^2 \delta_{m,n} + m l_B \delta_{m-1,n} \right), \\
&\langle \bar{\psi}_{m,k} | \partial_y \bar{\psi}_{n,k} \rangle = l_B \left( n \delta_{m-1,n} + (1/2) \delta_{m+1,n} \right), \\
&\langle \partial_x \bar{\psi}_{m,k} | \partial_x \bar{\psi}_{m,k} \rangle = (m+1/2) l_{B}^2+ l_B^4  k_{y}^2 , \\
&\langle \partial_x \bar{\psi}_{m,k} | \partial_y \bar{\psi}_{m,k} \rangle = i l_B^2/2, \\
&\langle \partial_y \bar{\psi}_{m,k} | \partial_y \bar{\psi}_{m,k} \rangle = l_B^2 (m+1/2).
\end{align}
The Hall conductivity is given as
\begin{align}
\sigma_{xy} &=\sum_m \int^{\pi/a_x}_{-\pi/a_x} d k_x \int^{\pi/a_y}_{-\pi/a_y} dk_y f( E^{LL}_{m} )  \mathrm{Re} \left[ \langle \partial_x \bar{\psi}_{m,k} | \partial_y \bar{\psi}_{m,k} \rangle-\langle \partial_y \bar{\psi}_{m,k} | \partial_x \bar{\psi}_{m,k} \rangle\right] \\
&= \sum_m f(E^{LL}_m),
\end{align}
where $a_x$ and $a_y$ represent the length of magnetic unit cell along $x$ direction and $y$ direction, respectively, which satisfy
\begin{align}
B a_x a_y =2 \pi.
\end{align}
This gives the same result in the main text. Note that for $d_\mathrm{max}=1$, we need to consider the degeneracy.
}

{
When we consider $r$ bands are occupied, the quantum geometric tensor takes the following form \cite{ozawa2021relations}:
\begin{align}
\chi_{ij}(\bm{k})=\sum^{r-1}_{m=0} \braket{\partial_i \bar{\psi}_{m,k}|1-P(\bm{k})|\partial_j\bar{\psi}_{m,k}},
\end{align}
where
\begin{align}
P(\bm{k})=\sum^{r-1}_{m=0} \ket{\bar{\psi}_{m,k}}\bra{\bar{\psi}_{m,k}}.
\end{align}
From a straightforward calculation, the Chern number, quantum metric, and Berry curvature for Landau levels are given by
\begin{align}
\mathcal{C} = -r,~~ g_{xx}=g_{yy}=\frac{r}{2}l_B^2, ~~g_{xy}=0,~~ \Omega_{xy}=-r l_B^2. \label{seq:geom}
\end{align}
These are the same results from Ref.~\cite{ozawa2021relations}.}

{
From this Chern number, the Hall conductivity is given by
\begin{align}
\sigma_{xy}= -C, \label{seq:Hall}
\end{align}
which is the same result above.
Note that for the case with $d_\mathrm{max} = 1$, $r$ in Eqs.~(\ref{seq:geom}) and (\ref{seq:Hall}) has to be replaced with $r-1$ due to the degeneracy at zero-energy.
}

{
As we see above, the geometry of the wavefunction is strongly related to $M'_0, \chi'_0$ and $\sigma_{xy}$, and they can be calculate from the wavefunctions. The same results with those obtained from Landau levels in the main text show that two approaches yield identical outcomes.}
\end{document}